\documentclass[12pt,journal,epsfig,onecolumn,draftclsnofoot]{IEEEtran}

\usepackage[dvips]{graphicx}
\usepackage{graphicx}
\usepackage{amssymb}
\usepackage{cite}
\usepackage{amsmath}
\usepackage{algorithm}
\usepackage{algorithmic}
\usepackage{multirow}
\usepackage[table]{xcolor}
\usepackage{subfigure}
  \usepackage{fancyhdr} 
\usepackage{makecell}
\usepackage{diagbox}
\usepackage{array}
\usepackage{threeparttable}
\usepackage{graphicx}
\graphicspath{ {./images/} }
\usepackage{caption,epstopdf}

\makeatother   

\newtheorem{lemma}{Lemma}

\begin{document}

\title{Target-Mounted Intelligent Reflecting Surface for Joint Location and Orientation Estimation
}


\author{Peilan Wang, Weidong Mei, ~\IEEEmembership{Member,~IEEE}, \\ Jun Fang, ~\IEEEmembership{Senior Member,~IEEE}, and Rui Zhang, ~\IEEEmembership{Fellow,~IEEE}
\thanks{Peilan Wang and Jun Fang are with the National Key Laboratory
of Science and Technology on Communications, University of
Electronic Science and Technology of China, Chengdu, China 611731 (e-mail: peilan\_wangle@std.uestc.edu.cn, JunFang@uestc.edu.cn).}
\thanks{Weidong Mei is with the Department of Electrical and Computer Engineering,
National University of Singapore, Singapore, 117583 (e-mail: wmei@nus.edu.sg).
}
\thanks{Rui Zhang is with the The Chinese University of Hong Kong, Shenzhen, and Shenzhen Research Institute of Big Data, Shenzhen, China, 518172 (e-mail: rzhang@cuhk.edu.cn). He is also with the Department of Electrical and Computer Engineering, National Uiversity of Singapore, Singapore 117583 (e-mail: elezhang@nus.edu.sg).}
}

\maketitle
\vspace{-12pt}

\begin{abstract}
Intelligent reflecting surface (IRS) has been widely recognized as an efficient technique to reconfigure the electromagnetic environment in favor of wireless communication performance. In this paper, we propose a new application of IRS for device-free target sensing via joint location and orientation estimation. In particular, different from the existing works that use IRS as an additional anchor node for localization/sensing, we consider mounting IRS on the sensing target, whereby estimating the IRS's location and orientation as that of the target by leveraging IRS's controllable signal reflection. To this end, we first propose a tensor-based method to acquire essential angle information between the IRS and the sensing transmitter as well as a set of distributed sensing receivers. Next, based on the estimated angle information, we formulate two optimization problems to estimate the location and orientation of the IRS/target, respectively, and obtain the locally optimal solutions to them by invoking two iterative algorithms, namely, gradient descent method and manifold optimization. In particular, we show that the orientation estimation problem admits a closed-form solution in a special case that usually holds in practice. Furthermore, theoretical analysis is conducted to draw essential insights into the proposed sensing system design and performance. Simulation results verify our theoretical analysis and demonstrate that the proposed methods can achieve high estimation accuracy which is close to the theoretical bound.


\end{abstract}

\begin{keywords}
Intelligent reflecting surface (IRS), device-free sensing, target-mounted IRS, location estimation, orientation estimation.
\end{keywords}

\section{Introduction}

In recent years, intelligent reflecting surface (IRS) or its equivalents such as reconfigurable intelligent surface (RIS) has been deemed as a promising technology for boosting the wireless communication spectral and energy efficiency cost-effectively, due to its capability of reshaping the wireless signal propagation environment with low-power and low-cost passive reflecting elements \cite{WuZhang20towards,RenzoZappone20,WuZhang21,ZhengYou22,PanZhou22}. Specifically, IRS is a planar surface consisting of a large number of passive elements, each of which can reflect the impinging wireless signal with a tunable phase shift and/or amplitude. By jointly adjusting the reflection of all its elements, IRS can alter the strength/direction of its reflected signal to achieve various useful functions such as passive beamforming, interference nulling/cancellation, spatial multiplexing, etc., for enhancing the wireless communication performance significantly \cite{WuZhang20towards,RenzoZappone20,WuZhang21,ZhengYou22,PanZhou22}.

On the other hand, recent studies have also revealed the great potential of IRS for enhancing the accuracy of wireless sensing and localization, by treating the IRS as an anchor node with a known location and leveraging its line-of-sight (LoS) link with the target for improving the sensing performance\cite{WangZhang21Joint,TengYuan22Bayes,He2020large,Elzanaty21,Buzzi22Foundation,Prasobh21,WangLiu22Tri}. Generally, wireless sensing can be classified into two categories, namely, device-based sensing and device-free sensing, where the targets need to and do not need to transmit/receive the sensing signals, respectively\cite{LiuHuang22}. For device-based sensing, the authors in \cite{WangZhang21Joint} formulated a least-square problem to localize the target based on the angles-of-departure (AoDs) from IRSs to it. Based on the message-passing algorithm, the authors in \cite{TengYuan22Bayes} proposed a Bayesian user localization and tracking algorithm to estimate and track the user position and derived the corresponding Bayesian Cram\'er-Rao bound (CRB). In addition, \cite{He2020large} proposed to leverage IRS to jointly estimate the location and two-dimensional (2D) orientation angles of a multi-antenna target by applying the maximum-likelihood estimation based on the time- and angles-of-arrival (AoAs) at the target, as well as other channel parameters. The results in \cite{He2020large} were further extended to a more general setup with three-dimensional (3D) orientation angle estimation for the target\cite{Elzanaty21}. While for device-free sensing, \cite{Buzzi22Foundation} considered utilizing IRSs to enhance the target detection performance in traditional MIMO radar systems, where the IRSs are placed in the vicinity of the radar transmitter/receiver to help illuminate the prospective passive targets. In addition, \cite{Prasobh21} introduced an IRS to enhance the sensing and communication capabilities of a dual-function radar, where its elements are adaptively partitioned to form multi-stage beams to localize the passive target while ensuring the communication quality of an intended user. Moreover, to achieve radar-like sensing capabilities in cellular networks, the authors in \cite{WangLiu22Tri} applied the trilateration method to localize multiple targets at the same time based on their distances to two base stations and one IRS with known locations. However, for the above IRS-assisted device-free sensing systems, their sensing performance relies heavily on the direct/IRS-reflected echo signals from the target to the receiver, which may be practically weak to achieve reliable sensing/localization due to the generally small radar cross section of the target and its random signal reflection. Moreover, the above works simply model each target as a point and thus are not applicable to estimate its orientation.


\begin{figure}[t]
\centering
\includegraphics[width=3.6in] {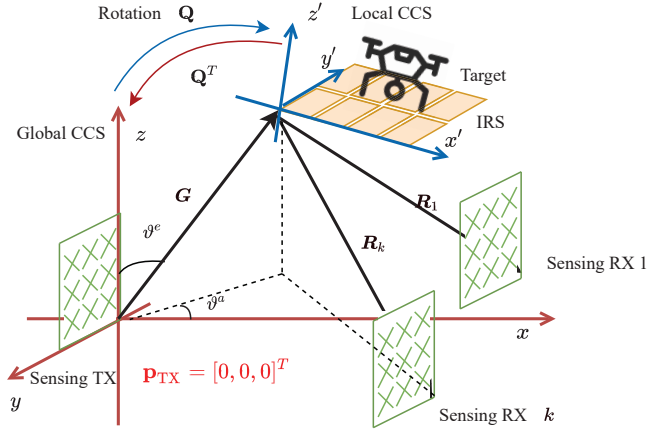}
\caption{Location and orientation estimation via a target-mounted IRS where the target can be e.g. a UAV over the air.}
\label{System_fig}
\end{figure}
\setlength{\textfloatsep}{0.2cm}  

In this paper, we propose a new device-free sensing system, where, instead of treating the IRS as an anchor node as in \cite{Buzzi22Foundation,Prasobh21,WangLiu22Tri}, we consider mounting it on the sensing target to facilitate its joint location and orientation estimation, as shown in Fig. \ref{System_fig}. By this means, we can equivalently estimate the IRS's location and orientation as that of the target and boost the reflected signal strength by leveraging IRS's controllable signal reflection and high spatial resolution thanks to its large aperture. To achieve this purpose, we first propose a new tensor-based method to obtain essential angle information between the IRS and the sensing transmitter (TX) as well as a set of distributed sensing receivers (RXs), by exploiting their cooperative beam searching. Then, based on the estimated angle information from the TX to the IRS and that from the IRS to all RXs, we show that the location estimation of the IRS/target can be formulated as a least-square problem, for which the gradient descent method is applied to obtain a locally optimal solution. Next, based on the estimated IRS's location and angle information from/to the TX/all RXs, we formulate another least-square problem to estimate its orientation and propose a manifold optimization method to solve it. In particular, it is shown that this problem admits a closed-form optimal solution in a special case which, fortunately, usually holds in practice. Furthermore, theoretical analysis is provided to obtain useful insights into the proposed sensing system design and performance. Specifically, we characterize the performance limits of the proposed estimation methods by deriving the CRBs on the involved angle information, as well as the IRS's location and orientation, respectively. In addition, we unveil several key factors affecting the accuracy of the proposed orientation estimation via the sensitivity analysis, such as the locations of the sensing TX and RXs, and the estimation accuracy of the IRS's angle information with the TX/RXs. Simulation results validate our theoretical analysis and demonstrate that the proposed methods can achieve high estimation accuracy close to the CRBs, thus providing an appealing solution to the location/orientation estimation problem in future wireless sensing systems.

The rest of this paper is organized as follows. Section \ref{sec-sys} presents the system model of the proposed IRS-mounted sensing system. Section \ref{sec-tensor} presents the proposed tensor-based method to estimate the required angle information for location and orientation estimation. Section \ref{sec-alg} presents the proposed algorithms for location and orientation estimation, respectively. Section \ref{sec-sens} presents the theoretical analysis on the proposed sensing scheme. Section \ref{sec-simulation} presents numerical results to verify the efficacy of our proposed methods. Finally, Section \ref{sec-conclu} concludes this paper.

{\it Notation:} In this paper, scalars, vectors, matrices and tensors are denoted by italic, bold-face lower-case, bold-face uppercase and bold-face calligraphic letters, respectively. For a matrix $\boldsymbol{A}$, its transpose, conjugate transpose, and determinant are denoted as $\boldsymbol{A}^T$, $\boldsymbol{A}^H$, and ${\rm det}(\boldsymbol{A})$, respectively. $\boldsymbol{I}_{M}$ denotes the identity matrix of size $M$. $\boldsymbol{C}^{x \times y}$ and $\boldsymbol{R}^{x \times y}$ denote the sets of $x \times y$-dimensional complex and real-valued matrices, respectively. $\| \cdot \|_F$ denotes the Frobenius norm of its argument. For a complex number $s$, $s^{\ast}$ and $|s|$ denote its conjugate and amplitude, respectively. $s \sim \mathcal{CN}(0,\sigma^2)$ means that $s$ is a circularly symmetric complex Gaussian (CSCG) random variable with zero mean and variance $\sigma^2$. $\mathbb{E}(\cdot) $ denotes the expected value of a random variable. $j$ denotes the imaginary unit, i.e., $j^2 =-1$. For a vector $\boldsymbol{a}$, ${\rm diag}(\boldsymbol{a})$ denotes a diagonal matrix whose diagonal elements are specified by $\boldsymbol{a}$. $\boldsymbol{1}_L$ denotes an $L$-dimensional vector with all the elements equal to $1$. $\odot$ denotes the Khatri-Rao product, while $\circ$ denotes the outer product. $\text{SO}(n)$ denotes the set of $n \times n$ orthogonal matrices with unit determinant, also known as special orthogonal (SO) group.


\section{System Model}
\label{sec-sys}
As shown in Fig. \ref{System_fig}, we consider a new IRS-assisted wireless sensing system with a sensing TX and a set of spatially distributed sensing RXs,\footnote{The TX and RXs can be, e.g., base stations (BSs) in cellular networks if a cellular-enabled sensing system is considered.} where an IRS is mounted on a mobile target (e.g. unmanned aerial vehicle (UAV) and unmanned ground vehicle (UGV)) to facilitate the sensing system to estimate its location and orientation information (which is assumed to be the same as that of the target) by exploiting its controllable signal reflection. We assume that the direct
links from the TX to all RXs have been estimated before the target location and orientation estimation by turning off the IRS's reflection; thus, each RX can retrieve the signal reflected by the IRS only when its reflection is turned on  by subtracting the signal received from the corresponding direct link from the received signal. We consider a narrowband system in this paper, while the results in this paper are applicable to the general broadband system by executing the proposed sensing scheme over any given frequency sub-band. For convenience, we establish a \emph{global Cartesian coordinate system (CCS)} in the considered sensing system. Due to the rotation of the IRS/target, to ease the computation of the angle information from the IRS (target) to the TX and all RXs, we also define a \emph{local CCS} at the IRS which lies in its $x'$-$y'$ plane, as shown in Fig. \ref{System_fig}.

\begingroup
\allowdisplaybreaks




\subsection{Geometric Model}
We consider that the TX is equipped with a uniform planar array (UPA) parallel to the $y$-$z$ plane of the global CCS, which comprises $N_t = N_{t_y} \times N_{t_z}$ antennas with $N_{t_y}$ and $N_{t_z}$ denoting the numbers of antennas along the $y$- and $z$-axes, respectively. Each distributed RX is equipped with a UPA comprising $N_r = {N_{r_y}} \times N_{r_z}$ antennas, with $N_{r_y}$ and $N_{r_z}$ denoting the numbers of antennas along the $y$- and $z$-axes, respectively. The IRS is a UPA consisting of $M = M_x \times M_y$ elements, where $M_x$ and $M_y$ denote the numbers of elements along the $x'$- and $y'$-axes of the local CCS, respectively. Without loss of generality, as depicted in Fig. \ref{System_fig}, we select the bottom-right elements of the UPAs at the TX and each RX $k$ as their reference elements, denoted as $\mathbf{p}_{\rm TX}=[p_{{\rm TX},x},p_{{\rm TX},y},p_{{\rm TX},z}]^T $ and $\mathbf{p}_{{\rm RX},k}=[p_{{\rm RX},k,x}, p_{{\rm RX},k,y},p_{{\rm RX},k,z}]^T ,k=1, 2,\ldots, K$, respectively, and the bottom-left element of that at the IRS as its reference element, denoted as $\mathbf{p}_{\rm I} = [p_{{\rm I},x}, p_{{\rm I},y},p_{{\rm I},z}]^T$. It is assumed that both $\mathbf{p}_{{\rm TX}}$ and each $\mathbf{p}_{{\rm RX},k}$ are known while $\mathbf{p}_{\rm I} $ is unknown, and its estimation will be studied in this paper. For convenience, we set $\mathbf{p}_{{\rm TX}}=[0,0,0]^T$, i.e., the TX is located at the origin. Moreover, let $\boldsymbol{\psi}=[\psi_z,\psi_y,\psi_x]^T$ represent the orientation of the IRS, where $\psi_z$, $\psi_y$, and $\psi_x$ are Euler angles denoting its degree of rotation around $z$-, $y$-, and $x$-axis, respectively. Note that the relationship between the local and global CCSs can be characterized by a $3\times 3$  rotation matrix in the 3D SO/rotation group, denoted as $\mathbf{Q} \in \text{SO}(3) \triangleq \{ \mathbf{Q} | {\rm det}(\mathbf{Q} )=1, \mathbf{Q}\mathbf{Q}^T = \boldsymbol{I} \}$. In particular, given the IRS/target's orientation $\boldsymbol{\psi}$, the rotation matrix $\mathbf{Q}$ can be expressed as \cite{Lynch17}
\begin{align}
\mathbf{Q} = \mathbf{Q}_z(\psi_z) \mathbf{Q}_y(\psi_y) \mathbf{Q}_x(\psi_x),
\label{eq-Qseq}
\end{align}
where $\mathbf{Q}_z(\psi_z)$ indicates the rotation of $\psi_z$ radians around the $z$-axis and is given by
\begin{align}
\mathbf{Q}_z(\psi_z) =\left[  \begin{matrix}
\cos(\psi_z) & -\sin(\psi_z) & 0 \\
\sin(\psi_z)  & \cos(\psi_z) & 0 \\
0 & 0 & 1
\end{matrix}
\right],
\end{align}
 $\mathbf{Q}_y(\psi_y)$ indicates the rotation of $\psi_y$ radians around the $y$-axis and is given by
 \begin{align}
\mathbf{Q}_y(\psi_y) =\left[  \begin{matrix}
\cos(\psi_y) & 0 & \sin(\psi_y)  \\
0 & 1 &0 \\
-\sin(\psi_y)  & 0 & \cos(\psi_y)  
\end{matrix}
\right],
\end{align}
and  $\mathbf{Q}_x(\psi_x)$ indicates the rotation of $\psi_x$ radians around the $x$-axis and is given by

\begin{align}
\mathbf{Q}_x(\psi_x) =\left[  \begin{matrix}
1 & 0 & 0 \\ 
0 & \cos(\psi_x) & -\sin(\psi_x)  \\
0 &\sin(\psi_x)  & \cos(\psi_x)  
\end{matrix}
\right].
\end{align}
Then, for any 3D location $^{\rm G}\mathbf{p} $ in the global CCS, its coordinates in the local CCS can be expressed as \cite{Lynch17} 
\begin{align}
^{\rm L}\mathbf{p} = \mathbf{Q}^T  \big( ^{\rm G}\mathbf{p}  - \mathbf{p}_{\rm I} \big).
\label{eq-Transform}
\end{align}
Obviously, we have $^{\rm L} \mathbf{p}_{\rm I}=[0,0,0]^T$, i.e., the reference element of the IRS is at the origin of the local CCS. It follows that the IRS's orientation with respect to (w.r.t.) the global CCS can be characterized by the rotation matrix $\mathbf{Q}$. As such, in this paper, we aim to jointly estimate the IRS's location and rotation matrix, denoted as $\mathcal{Q} \triangleq \{\mathbf{p}_{\rm I}, \mathbf{Q} \}$, which has six unknowns (dimensions) in total. In the sequel of this paper, we refer to the estimation of $\mathcal{Q}$ as \emph{six-dimensional (6D) information acquisition}.


\subsection{Channel Model}
Let $\boldsymbol{G} \in \mathbb C^{M \times N_t}$ and $\boldsymbol{R}_k \in \mathbb C^{N_r \times M}, k=1,2, \ldots, K$ denote the TX-IRS and IRS-RX $k $ channels, respectively. In this paper, we assume that the above channels are dominated by LoS propagation, which generally holds in practice for high operating frequency (e.g., millimeter-wave/terahertz (mmWave/THz) \cite{WangFang21a}) and/or high-altitude target (e.g. UAV \cite{MeiZhang21}).\footnote{As will be shown in Section IV via simulation, the proposed 6D information acquisition method is still applicable if there exist non-LoS components in the involved channels, as long as the strength of their LoS components is sufficiently large.} To characterize the LoS channels above, we note that the TX-IRS and IRS-RX $k$ distances are much larger than the size of UPAs at the TX, IRS,  and each RX $k$. Thus, the signals from/to the source/destination nodes can be approximated as uniform planar waves at each UPA, and the LoS channel between the IRS and the TX/RX $k$ can be modeled as the outer product of steering vectors at the two sides.

Specifically, let $\theta^e$/$\theta^a$ denote the elevation/azimuth AoA or AoD at a UPA. Then, for the UPA parallel to the $y$-$z$ plane, its steering vector is given by\cite{Van04}
\begin{align}
\boldsymbol{a}( \theta^e,\theta^a )  =& [1, \ldots, e^{j \frac{2\pi d} {\lambda} (N_y-1) \sin(\theta^e) \sin(\theta^a)  }] ^T\otimes  [1, \ldots, e^{j \frac{2\pi d} {\lambda} (N_z-1) \cos(\theta^e)   }] ^T ,
\label{steer-yz}
\end{align}where $d$ is the spacing between any two adjacent antennas/elements on the UPA, $\lambda$ denotes the wavelength, $N_y$ and $N_z$ denote the numbers of antennas/elements along the $y$- and $z$-axes, respectively.
While for the UPA parallel to the $x$-$y$ plane, its steering vector is given by \cite{Van04}
\begin{align}
\boldsymbol{a}(\theta^e,\theta^a) = & [1, \ldots, e^{j \frac{2 \pi d}{\lambda} (N_x-1)\sin (\theta^e) \cos (\theta^a) }]^T   \otimes [1, \ldots, e^{j \frac{2\pi d}{\lambda} (N_y-1)\sin(\theta^e) \sin(\theta^a)}]^T ,
\label{steer-xy}
\end{align}where $N_x$ denotes the number of antennas/elements along the $x$-axis.

Then, for the UPA at the TX, let $\vartheta^e$/$\vartheta^a$ denote its elevation/azimuth AoD w.r.t. the IRS, which, based on the geometry in Fig. \ref{System_fig}, is given by
\begin{align}
\vartheta^e = \arccos \frac{{p}_{{\rm I},z} - {p}_{{\rm TX},z} }{ \| \mathbf{p}_{{\rm I}} - \mathbf{p}_{{\rm TX}} \|_2},  \quad \vartheta^a  = \arctan\frac{{p}_{{\rm I},y} - {p}_{{\rm TX},y} }{ {p}_{{\rm I},x} - {p}_{{\rm TX},x}}.
\label{eq-AoD}
\end{align}
As the Tx's UPA lies in the $y$-$z$ plane, it follows from \eqref{steer-yz} that its steering vector is given by
\begin{align}
\boldsymbol{a}_t(\zeta_0^e,\zeta_0^a )  =& [1, \ldots, e^{j \pi  (N_{t_y}-1)\zeta_0^a  }] ^T \otimes  [1, \ldots, e^{j \pi (N_{t_z}-1) \zeta_0^e   }] ^T , 
\label{zeta0}
\end{align}
where $\zeta_0^a \triangleq \frac{2d_t}{\lambda} \sin(\vartheta^e) \sin(\vartheta^a)$ and $\zeta_0^e \triangleq \frac{2d_t}{\lambda} \cos(\vartheta^e)$ are defined as the effective azimuth and elevation spatial frequencies at the TX, respectively, and $d_t$ denotes the antenna spacing of its UPA.
Similarly, for the UPA at RX $k$, let $\gamma^e_k$/$\gamma^a_k$ denote its elevation/azimuth AoA w.r.t. the IRS, which is given by
\begin{align}
\gamma_k^e = \arccos \frac{p_{{\rm RX},k,z}  -  {p}_{{\rm I},z}   }{ \| \mathbf{p}_{{\rm RX},k} - \mathbf{p}_{\rm I}\|_2},\quad \gamma_k^a = \arctan \frac{  p_{{\rm RX},k,y}  -  {p}_{{\rm I},y}    }{  p_{{\rm RX},k,x}  -  {p}_{{\rm I},x}   }.
\label{eq-AoA}
\end{align}
Then, the steering vector of its UPA is written as
\begin{align}
\boldsymbol{a}_{r} ({\zeta}_k^e,\zeta_k^a) = &  [1, \ldots, e^{j \pi  (N_{r_y}-1) \zeta_k^a}]^T   \otimes [1, \ldots, e^{j \frac{2\pi d_r}{\lambda} (N_{r_z}-1)\zeta_k^e}]^T ,
\label{zetak}
\end{align}
where $\zeta_k^a \triangleq \frac{2d_r}{\lambda}\sin(\gamma^e_k) \sin(\gamma^a_k)$ and $\zeta_k^e \triangleq \frac{2d_r}{\lambda} \cos(\gamma_k^e)$ denote the effective spatial azimuth and elevation frequencies at RX $k$, respectively, and $d_r$ denotes the antenna spacing of the UPA at RX $k$.

Nevertheless, deriving the AoD and AoA at the IRS is more complicated due to its 3D rotation, which requires the transformation between the global and local CCSs. Specifically, based on \eqref{eq-Transform}, let $^{\rm L} \mathbf{p}_{\rm TX} = \mathbf{Q}^T ( \mathbf{p}_{\rm TX} - \mathbf{p}_{\rm I})$ denote the TX's coordinates in the local CCS, and
\begin{align}
\mathbf{q}_A \triangleq   ^{\rm L}\mathbf{p}_{\rm I}- ^{\rm L} \mathbf{p}_{\rm TX} =- \mathbf{Q}^T  ( \mathbf{p}_{\rm TX} - \mathbf{p}_{\rm I}) = [q_{A,x},q_{A,y},q_{A,z}]^T
\label{qA}
\end{align}denotes the direction vector from the TX to the IRS in the local CCS.

Let $\phi_A^e$/$\phi_A^a$ denote the elevation/azimuth AoA at the IRS w.r.t. the TX. Now based on \eqref{qA}, $\phi_A^e$ and $\phi_A^a$ can be efficiently calculated as
\begin{align}
\phi_{A}^e =&  \arccos \frac{q_{A,z}}{\| \mathbf{q}_{A}\|}  , \quad
\phi_{A}^a = {\rm arctan} \frac{q_{A,y}}{q_{A,x}},
\label{qA-ang}
\end{align}
respectively.
As the IRS lies in the $x'$-$y'$ plane of the local CCS, based on \eqref{steer-xy}, its receive steering vector w.r.t. the TX is given by
\begin{align}
\boldsymbol{a}_I(\omega_A^e, \omega_A^a) = & [1, \ldots, e^{j \pi (M_x-1)\omega_A^a}]^T   \otimes [1, \ldots,  e^{j \pi (M_y-1)\omega_A^e}]^T    ,
\label{steer-R}
\end{align}where $\omega_A^a \triangleq \frac{2  d_{I}}{\lambda}\sin (\phi_A^e) \cos (\phi_A^a)$ and $\omega_A^e \triangleq \frac{2d_I}{\lambda}  \sin(\phi_A^e) \sin(\phi_A^a)$ are effective receive azimuth and elevation spatial frequencies at the IRS, respectively, and $d_I$ denotes the element spacing of the IRS.

Similarly, let $\phi_{D_k}^e$/$\phi_{D_k}^a$ denote the elevation/azimuth AoD from the IRS to RX $k$, which is given by
\begin{align}
\phi_{D_k}^e =&  \arccos \frac{q_{{D_k},z}}{\| \mathbf{q}_{D_k}\|}  , \quad
\phi_{D_k}^a = {\rm arctan} \frac{q_{{D_k},y}}{q_{{D_k},x}},
\label{qDk-ang}
\end{align}
with
\begin{align}
\mathbf{q}_{D_k} &\triangleq\mathbf{Q}^T( \mathbf{p}_{{\rm RX},k}- \mathbf{p}_{\rm I}  ) =  [{q}_{D_k,x},{q}_{D_k,y},{q}_{D_k,z}]^T .
\label{q_Dk}
\end{align}
Then, the IRS's transmit steering vector w.r.t. RX $k$ can be obtained by replacing $\omega_{A}^a$ and $\omega_{A}^{e}$ in \eqref{steer-R} with $ \omega_{D_k}^a \triangleq \frac{2d_I}{\lambda}\sin (\phi_{D_k}^e) \cos (\phi_{D_k}^a)$ and $ \omega_{D_k}^e \triangleq \frac{2d_I}{\lambda}\sin (\phi_{D_k}^e) \sin (\phi_{D_k}^a)$, respectively, which denote the effective transmit azimuth and elevation spatial frequencies at the IRS.

As a result, the BS-IRS channel $\boldsymbol{G} \in \mathbb C^{M \times N_t}$ and IRS-RX $k$ channel $\boldsymbol{R}_k \in \mathbb C^{N_r \times M}$ can be written as
\begin{align}
\boldsymbol{G} = &  \alpha_G \boldsymbol{a}_I(\omega_A^e, \omega_A^a)\boldsymbol{a}_t^H (\zeta_0^e,\zeta_0^a )  , \quad \boldsymbol{R}_k = \alpha_{r_k}\boldsymbol{a}_{r} ({\zeta}_k^e,\zeta_k^a)  \boldsymbol{a}_I^H(\omega_{D_k}^e, \omega_{D_k}^a), k=1,2, \ldots, K,
\label{LOS-1}
\end{align}
where $\alpha_G$  and $\alpha_{r_k}$ denote the complex channel coefficients, which subsume the BS/IRS antenna/element gain and path loss.


\subsection{Signal Model}
To compensate for the double path loss incurred by the IRS, the TX and each RX should be equipped with a sufficiently large number of antennas. To reduce the resulting hardware cost, we consider that analog transmit and receive beamforming are applied at  the TX and all RXs, respectively. Let $\boldsymbol{f} =[f_1,f_2,\cdots,f_{N_t}]^H \in \mathbb C^{N_t \times 1}$ and $s$ denote the analog transmit beamforming vector and the transmitted symbol, respectively, with $\lvert f_i \rvert = 1/\sqrt{N_t}, \forall i$ and $ \mathbb E[\lvert s \rvert^2]=1$. Then, the transmitted signal is expressed as $\boldsymbol{x}= \sqrt{P_t}\boldsymbol{f}{s}$,
where $P_t$ denotes the transmit power. Moreover, let $\boldsymbol{\Phi} = {\rm diag} (\boldsymbol{v}^H )$  be the reflection coefficient matrix of the IRS with $\boldsymbol{v}\triangleq [ e^{j \theta_1 }, \ldots, e^{j \theta_M }] ^H \in \mathbb C^{M\times 1}$ denoting its passive beamforming vector and $\boldsymbol{w}_k =[w_{k,1},w_{k,2},\cdots,w_{k,N_r}]^H \in \mathbb C^{N_{r} \times 1}$ be the analog receive beamforming vector at each RX $k$, with $\lvert w_{k,i} \rvert = 1/\sqrt{N_r}, \forall k,i$. Then, the received signal at RX $k$ after antenna combining is given by
\begin{align}
y_k = & \sqrt{P_t}\boldsymbol{w}_k^H \boldsymbol{R}_k \boldsymbol{\Phi} \boldsymbol{G} \boldsymbol{f} {s} + \boldsymbol{w}_k^H \boldsymbol{n}_k,
\end{align}where $\boldsymbol{n}_k \in \mathbb C^{N_r \times 1} \sim \mathcal{CN}(0,\sigma^2 \boldsymbol{I}_{N_r} )$ denotes the background noise, with $\sigma^2$ denoting the noise power per antenna.

For the considered narrowband system, we can use angle information \emph{only} for 6D information acquisition. In the sequel, we will firstly introduce how to estimate the angle information required by formulating a tensor decomposition-based problem. Next, we will elaborate the proposed angle-based method to estimate the IRS's location and orientation jointly.

\begin{figure}[t]
\centering
\includegraphics[width=5in] {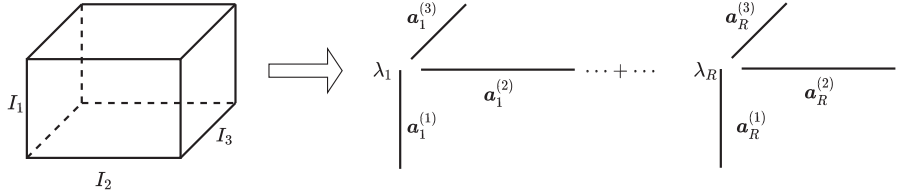}
\caption{Illustration of CP decomposition.}
\label{Fig-tensor}
\end{figure}

\section{Tensor-Based Angle Information Acquisition}
\label{sec-tensor}
Tensor decomposition has been shown as an effective tool for angle estimation in wireless communications/sensing \cite{ZhouFang17}. In particular, the reflected link via IRS makes it efficient to construct a tensor-based received signal model thanks to the two-hop signal structure. In this section, we formulate a tensor decomposition-based problem and propose an efficient algorithm to acquire the associated angle information.

\subsection{Preliminary of Tensors}
First, we provide a brief introduction on tensors. Interested reader can refer to \cite{Kolda09tensor} for a detailed overview of tensors. Basically, a tensor generalizes vectors and matrices to higher dimensions (ways/modes), and vectors and matrices can be viewed as a special case of tensors with one and two modes, respectively.
For ease of exposition, we take a third-order tensor as an example in this subsection, which is denoted as $\mathcal{\boldsymbol{\chi }} \in \mathbb C^{I_1 \times I_2 \times  I_{3}}$  and depicted in Fig. \ref{Fig-tensor}, with its $(i_1,i_2,i_3)$-th entry denoted by ${\chi}[{i_1,i_2,i_3}]$. For a tensor, its subarray with all but one index fixed is termed a {\it fiber}, while that with all but two indices fixed is termed a {\it slice}. As such, a fiber is analogous to a row/column in a matrix, while a slice is equivalent to a matrix. For example, the $i_3$-th frontal slice of $\boldsymbol{\chi}$ is a matrix $\boldsymbol{\chi}_{:,:,i_3} \in \mathbb C^{I_1 \times I_2}$ by fixing its third dimension as $i_3$. Moreover, the mode-$n$ fiber refers to the vector obtained by fixing all indices of a tensor except for its $n$-th dimension. \emph{Unfolding} is an operation turning a tensor into a matrix. The mode-$n$ unfolding of a tensor $\boldsymbol{\chi}$, denoted by $\boldsymbol{\chi}_{(n)}$, arranges all of its mode-$n$ fibers to be the columns of a matrix. 

The CANDECOMP/PARAFAC (CP) decomposition decomposes the tensor $\boldsymbol{\chi}$ into a sum of rank-one tensor components as depicted in Fig. \ref{Fig-tensor}, i.e.,
\begin{align}
\mathcal{\boldsymbol{\chi}} = \sum_{r=1}^R \lambda_r \boldsymbol{a}_r^{(1)} \circ \boldsymbol{a}_r^{(2)} \circ \boldsymbol{a}_r^{(3)}, \label{eq-t3}
\end{align}
where $\boldsymbol{a}_r^{(n)} =[{a}_{r,1}^{(n)}, \ldots, {a}_{r,I_n}^{(n)}]^T \in \mathbb C^{I_n \times 1},n=1,2,3$, and the minimum $R$ that satisfies \eqref{eq-t3} is referred to as the rank of the tensor $\mathcal{\boldsymbol{\chi}}$. Based on \eqref{eq-t3}, we have
\begin{align}
\chi [i_1,i_2,i_3] = \sum_{r=1}^R \lambda_r  {a}_{r,i_1}^{(1)} {a}_{r,i_2}^{(2)} {a}_{r,i_3}^{(3)}.
\end{align}
Define $\boldsymbol{A}^{(n)} \triangleq [ \boldsymbol{a}_1^{(n)}, \ldots, \boldsymbol{a}_R^{(n)}] \in \mathbb C^{I_n \times R}$ as the factor matrix of $\boldsymbol{\chi}$ along the $n$-th mode. Then, we can express the mode-$n$ ($n=1,2,3$) unfolding of $\boldsymbol{\mathcal{\chi}}$ as
\begin{align}
\boldsymbol{\chi}_{(1)} = \boldsymbol{A}^{(1)}  \boldsymbol{{\Lambda}}( \boldsymbol{A}^{(3)} \odot \boldsymbol{A}^{(2)})^T , \label{eq-mode1} \\
\boldsymbol{\chi}_{(2)} = \boldsymbol{A}^{(2)}  \boldsymbol{{\Lambda}}( \boldsymbol{A}^{(3)} \odot \boldsymbol{A}^{(1)})^T , \\
\boldsymbol{\chi}_{(3)} = \boldsymbol{A}^{(3)}  \boldsymbol{{\Lambda}}( \boldsymbol{A}^{(2)} \odot \boldsymbol{A}^{(1)})^T , \label{eq-mode3}
\end{align}
where $\boldsymbol{\Lambda} \triangleq {\rm diag}(\lambda_1, \ldots, \lambda_R)$.

\subsection{Tensor-Based Formulation}
To facilitate the practical implementation, we consider that each RX $k$, the TX, and the IRS employ predefined codebooks for 6D information acquisition, denoted as $\{ \boldsymbol{w}_{k,i}\}_{i=1}^{D_{\rm RX}}$, $\{ \boldsymbol{f}_{i} \}_{j=1}^{D_{\rm TX}} $, and $\{ \boldsymbol{v}_{q} \}_{q=1}^{D_{\rm IRS}}$, respectively, where $D_{\rm RX}$, $D_{\rm TX}$, and $D_{\rm IRS}$ are the numbers of corresponding beam codewords. Accordingly, let
\begin{align}
\boldsymbol{W}_k \triangleq  & [\boldsymbol{w}_{k,1}, \ldots,  \boldsymbol{w}_{k,D_{\rm RX}}] \in \mathbb C^{N_{r} \times D_{\rm RX}} , 
\boldsymbol{F} \triangleq   [\boldsymbol{f}_{1}, \ldots,  \boldsymbol{f}_{D_{\rm TX}}] \in \mathbb C^{N_{t} \times D_{\rm TX}},  \nonumber \\
\boldsymbol{V} \triangleq  & [\boldsymbol{v}_{1}, \ldots,  \boldsymbol{v}_{D_{\rm IRS}}] \in \mathbb C^{M\times D_{\rm IRS}}, \label{eq-V}
\end{align}
denote the codebook matrices at RX $k$, the TX, and the IRS, respectively.
The TX, IRS, and all RXs consequently tune the beamforming vectors based on their respective codebooks over time. Then, with different beam combinations at the TX, IRS, and RXs, the received signal at each RX can be expressed as a third-order tensor,  $\boldsymbol{\mathcal{Y}}_k \in \mathbb C^{D_{\rm RX} \times D_{\rm TX} \times D_{\rm IRS}}$, with its $(i,j,q)$-th entry denoting the received signal at RX $k$ when it applies the codeword $\boldsymbol{w}_{k,i}$, and the IRS and TX apply the codewords $\boldsymbol{v}_q$ and $\boldsymbol{f}_j$, respectively, i.e.,
\begin{align}
\mathcal{Y}_k[i,j,q] = & \sqrt{P_t} \boldsymbol{w}_{k,i}^H \boldsymbol{R} \boldsymbol{\Phi}_q \boldsymbol{G} \boldsymbol{f}_j + n_{k,i,j,q}, \nonumber \\
=&   \sqrt{P_t}\alpha_k \big( \boldsymbol{w}_{k,i}^H  \boldsymbol{a}_{r} ({\zeta}_k^e,\zeta_k^a)  \big)  \times \big( \boldsymbol{a}_t^H(\zeta_0^e,\zeta_0^a )\boldsymbol{f}_j \big) \times \big(   \boldsymbol{v}^H _q   \boldsymbol{ a}_I (\eta_k^e, \eta_k^a )  \big)  + n_{k,i,j,q},
\label{eq-eta}
\end{align}
where $\boldsymbol{\Phi}_q = {\rm diag} (\boldsymbol{v}_q)$, $n_{k,i,j,q} \sim \mathcal{CN}(0,\sigma^2)$ is the effective received noise at RX $k$ after its combining,  $ \alpha_k \triangleq \alpha_G\alpha_{r_k}$ denotes the end-to-end path gain from the TX to RX $k$, and $\eta_k^e =\omega_A^e - \omega_{D_k}^e$ and $\eta_k^a=\omega_A^a - \omega_{D_k}^a$ are referred to as the cascaded elevation and azimuth spatial frequencies at the IRS w.r.t. RX $k$, respectively.

Next, we focus on the $q$-th frontal slice of $\boldsymbol{\mathcal{Y}}_k$, i.e., the IRS adopts the codeword $\boldsymbol{v}_q$ while the TX and RX $k$ conduct beam searching with the codebooks $\boldsymbol{F}$ and $\boldsymbol{W}_k$, respectively. In this case, the received signal at RX $k$ can be expressed as
\begin{align}
\boldsymbol{Y}_{k,q} = & \boldsymbol{W}_k^H \boldsymbol{R}_k \boldsymbol{\Phi}_q\boldsymbol{G} \boldsymbol{F} +  \boldsymbol{N}_{k,q}
\stackrel{(a)} =  \boldsymbol{\tilde a}_{r,k}  \boldsymbol{\tilde a}_{t} ^T \times  \tilde a_{I,k,q} +   \boldsymbol{N}_{k,q}, 
\label{eq-rec-q}
\end{align}
where $\boldsymbol{N}_{k,q}\in \mathbb C^{D_{\rm RX} \times D_{\rm TX}}$ is the stacked received noise matrix with $n_{k,i,j,q}$ being its $(i,j)$-th entry, and in $(a)$ we define
\begin{align}
\boldsymbol{\tilde a}_{r,k}   \triangleq & \boldsymbol{W}_k^H\boldsymbol{a}_{r} ({\zeta}_k^e,\zeta_k^a) ,  \;\;
\boldsymbol{\tilde{a}}_{t}   \triangleq  \boldsymbol{F}^{T} \boldsymbol{a}_t^{\ast} (\zeta_0^e,\zeta_0^a ), \;\; 
\tilde{a}_{I,k,q} \triangleq  \sqrt{P_t}\alpha_k \boldsymbol{v}^H_{q}   \boldsymbol{ a}_I (\eta_k^e, \eta_k^a ) . 
\end{align}
It follows from \eqref{eq-rec-q} that $\boldsymbol{Y}_{k,q} $ can be represented as the rank-one outer product of two vectors $\boldsymbol{\tilde a}_{r,k}$ and $\boldsymbol{\tilde a}_t$. Hence, the tensor $\boldsymbol{\mathcal{Y}}_k$ admits the following CP decomposition,
\begin{align}
\boldsymbol{\mathcal{Y}}_k = \boldsymbol{\tilde a}_{r,k} \circ \boldsymbol{\tilde a}_t \circ \boldsymbol{\tilde a}_{I,k} + \boldsymbol{\mathcal{N}}_k,
\label{eq-Yk}
\end{align}
where $\boldsymbol{\mathcal{N}}_k$ is the associated received noise tensor with $\boldsymbol{N}_{k,q}$ being its $q$-th frontal slice, and
\begin{align}
\boldsymbol{\tilde a}_{I,k} \triangleq  &\sqrt{P_t}\alpha_k \boldsymbol{V}^H  \boldsymbol{ a}_I (\eta_k^e, \eta_k^a )= [\tilde{a}_{I,k,1}, \ldots, \tilde{a}_{I,k,q}, \ldots, \tilde{a}_{I,k,D_{\rm IRS}}]^T.
\label{tilde-aI}
\end{align}
Following \eqref{eq-mode1}-\eqref{eq-mode3}, the mode-$n$ ($n=1,2,3$) unfoldings of tensor $\boldsymbol{\mathcal{Y}}_k$ in \eqref{eq-Yk} can be expressed as 
\begin{align}
\boldsymbol{Y}_{k,(1)}=  \boldsymbol{\tilde a}_{r,k} \big(  \boldsymbol{\tilde a}_{I,k}  \odot \boldsymbol{\tilde a}_t   \big)^ T + \boldsymbol{N}_{k,(1)}, \label{eq-Ym1}\\
\boldsymbol{Y}_{k,(2)} =  \boldsymbol{\tilde a}_{t} \big(  \boldsymbol{\tilde a}_{I,k}  \odot \boldsymbol{\tilde a}_{r,k}   \big)^ T + \boldsymbol{N}_{k,(2)}, \\
\boldsymbol{Y}_{k,(3)} =  \boldsymbol{\tilde a}_{I,k} \big(  \boldsymbol{\tilde a}_{t}  \odot \boldsymbol{\tilde a}_{r,k}   \big)^ T + \boldsymbol{N}_{k,(3)}, \label{eq-Ym3}
\end{align}
where $\boldsymbol{N}_{k,({n})}$ denotes the mode-$n$ unfolding of the noise tensor $\boldsymbol{\mathcal{N}}_k$. 
It should be mentioned that in the proposed tensor decomposition-based method, the codebook matrices $\boldsymbol{W}_k$, $\boldsymbol{F}$, and $\boldsymbol{V}$ in \eqref{eq-V} can be generated by randomly selecting the phase shifts within $[0,2\pi]$. As will be shown in Section \ref{sec-simulation} via simulation, the required number of beam combinations by this means, i.e., $ D_{\rm RX}D_{\rm TX}D_{\rm IRS}$, can be much smaller than that of exhaustive beam searching within the overall angle domain, i.e., $N_tN_rM$.

\subsection{Factor Vectors Estimation}
It is noted that the required angle parameters, i.e., $\{ \zeta_k^e, \zeta_k^a \}$, $\{\zeta_0^e, \zeta_0^a\}$, and $\{\eta_k^e,\eta_k^a\}$, are decoupled in the three factor vectors, $\boldsymbol{\tilde a}_{r,k}$, $\boldsymbol{\tilde a}_{t}$, and $\boldsymbol{\tilde a}_{I,k}$, respectively. As such, we first estimate these three factor vectors, from which the angle information can be extracted. Specifically, based on \eqref{eq-Yk}, we can estimate the three factor vectors $\{ \boldsymbol{\tilde a}_{r,k} ,\boldsymbol{\tilde a}_t , \boldsymbol{\tilde a}_{I,k}  \}$ by solving the following least-square problem,
\begin{align}
\min_{ \{ \boldsymbol{\tilde a}_{r,k},\boldsymbol{\tilde a}_{t},\boldsymbol{\tilde a}_{I,k} \} } \| \boldsymbol{\mathcal{Y}}_k - \boldsymbol{\tilde a}_{r,k} \circ \boldsymbol{\tilde a}_t \circ \boldsymbol{\tilde a}_{I,k} \|_F^2,
\label{eq-opt-Y}
\end{align}for which the classical alternating least-squares (ALS) method \cite{Kolda09tensor,Cichocki15tensor} can be applied, where the three factor vectors are alternately optimized with two of them fixed. Let $\boldsymbol{\hat a}_{i}^{(l)}$ denote the optimized factor vector of $\boldsymbol{\tilde a}_{i}, i\in \{  \{r,k \}, \{I,k \},t\} $ in the $l$-th iteration of the ALS method Then, the iteration can proceed as 
\begin{align}
\boldsymbol{\hat a}_{r,k}^{(l+1)} = &\arg \min_{ \boldsymbol{\tilde a}_{r,k}    } \|  \boldsymbol{Y}_{k,(1)} -  \boldsymbol{\tilde a}_{r,k} \big(  \boldsymbol{\hat a}_{I,k}^{(l)}  \odot \boldsymbol{\hat a}_t ^{(l)}  \big)^ T \|_F^2  \label{eq-Yk1}, \\
\boldsymbol{\hat a}_{t}^{(l+1)} =& \arg \min_{ \boldsymbol{\tilde a}_{t}    } \|  \boldsymbol{Y}_{k,(2)} -  \boldsymbol{\tilde a}_{t} \big(  \boldsymbol{\hat a}_{I,k}^{(l)}  \odot \boldsymbol{\hat a}_{r,k}^{(l+1)}  \big)^ T \|_F^2, \\
\boldsymbol{\hat a}_{I,k}^{(l+1)} =& \arg \min_{ \boldsymbol{\tilde a}_{I,k}    } \|  \boldsymbol{Y}_{k,(3)} -  \boldsymbol{\tilde a}_{I,k} \big(  \boldsymbol{\hat a}_{t}^{(l+1)}  \odot \boldsymbol{\hat a}_{r,k}^{(l+1)}  \big)^ T \|_F^2 \label{eq-Yk3}.
\end{align}
The least-square problems in \eqref{eq-Yk1}-\eqref{eq-Yk3}  admit closed-form solutions, which are given by 
$\boldsymbol{\hat a}_{r,k}^{(l+1)}  =   \boldsymbol{Y}_{k,(1)} \big((  \boldsymbol{\hat a}_{I,k}^{(l)}  \odot \boldsymbol{\hat a}_t ^{(l)}  )^ T \big)^{\dagger}$, $\boldsymbol{\hat a}_{t}^{(l+1)} =  \boldsymbol{Y}_{k,(2)} \big( (  \boldsymbol{\hat a}_{I,k}^{(l)}  \odot \boldsymbol{\hat a}_{r,k}^{(l+1)})^T  \big)^ {\dagger}$, and $\boldsymbol{\hat a}_{I,k}^{(l+1)} = \boldsymbol{Y}_{k,(3)} \big( ( \boldsymbol{\hat a}_{t}^{(l+1)}  \odot \boldsymbol{\hat a}_{r,k}^{(l+1)} )^T\big)^ {\dagger}$, respectively.
The ALS iteration can proceed until the objective value of \eqref{eq-opt-Y} is below a predefined threshold.

\subsection{Angle Information Acquisition}
Let $\boldsymbol{\hat a}_{r,k}$, $\boldsymbol{\hat a}_{t}$, and $\boldsymbol{\hat a}_{I}$ denote the estimates of the factor vectors $\boldsymbol{\tilde a}_{r,k}$, $\boldsymbol{\tilde a}_{t}$, and $\boldsymbol{\tilde a}_{I,k}$ by the ALS method, respectively. Next, we apply a correlation-based method to estimate the associated angle parameters from them. Note that this method can be viewed as a maximum-likelihood estimator \cite{ZhouFang17}, and thus their resulting estimation errors follow independent and identically distributed (i.i.d.) CSCG distribution.

First, we can estimate the effective spatial elevation and azimuth frequencies at RX $k$ as
\begin{align}
\{ \hat{\zeta}_{k}^e, \hat{\zeta}_{k}^a \}   = & \arg \max_{ \{ \zeta_k^e,\zeta_k^a \}  } \left \{  \frac{ |\boldsymbol{\hat a}_{r,k} ^H \boldsymbol{W}_k^H \boldsymbol{ a}_{r} (\zeta_k^e,\zeta_k^a) | }{ \| \boldsymbol{\hat a}_{r,k} \|_2  \|   \boldsymbol{W}_k^H \boldsymbol{ a}_{r} (\zeta_k^e,\zeta_k^a)\| _2    }  \right \} .
\label{eq-zeta-k}
\end{align}
Similarly, the cascaded spatial elevation and azimuth frequencies at the IRS and those at the TX are respectively estimated as
\begin{align}
\{ \hat{\eta}_{k}^e, \hat{\eta}_{k}^a \}   =&  \arg \max_{ \{ \eta_k^e,\eta_0^a \}  } \left \{  \frac{ |\boldsymbol{\hat a}_{I,k} ^H  \boldsymbol{V}^H \boldsymbol{ a}_{I}(\eta_k^e,\eta_k^a) | }{ \| \boldsymbol{\hat a}_{I,k} \|_2  \|  \boldsymbol{V}^H  \boldsymbol{a}_I(\eta_0^e,\eta_0^a) \| _2    }  \right \} , \\
\{ \hat{\zeta}_{0}^e, \hat{\zeta}_{0}^a \}   =& \arg \max_{ \{ \zeta_0^e,\zeta_0^a \}  } \left \{  \frac{ |\boldsymbol{\hat a}_{t} ^H \boldsymbol{F}^T \boldsymbol{ a}_t^{\ast} (\zeta_0^e,\zeta_0^a) | }{ \| \boldsymbol{\hat a}_{t} \|_2  \|   \boldsymbol{F}^T \boldsymbol{a}_t^{\ast} (\zeta_0^e,\zeta_0^a) \| _2    }  \right \} .
\label{eq-zeta0}
\end{align}

However, a two-dimensional search is needed to solve \eqref{eq-zeta-k}-\eqref{eq-zeta0}, which may result in a high searching complexity. To accelerate the search, we can first employ a coarse grid search within $[-1,1]$ and then gradually refine the search in the vicinity of the possible grids \cite{ZhouFang17}.
After the above estimations, the estimated angle parameters are fed back from the RXs to the TX which then estimates the IRS's location and orientation, as will be detailed in the following section.

\section{Proposed Algorithm for 6D Information Acquisition}
\label{sec-alg}
In this section, we discuss how to acquire the 6D information from the obtained angle information. In particular, it is noted from \eqref{qA-ang}, \eqref{qDk-ang}, and \eqref{eq-eta} that the angle parameters $\{  {\eta}_k^a, \eta_k^e\}_{k=1}^K$ are involved in both the IRS's location information $\mathbf{p}_{\rm I}$ and the rotation matrix $\mathbf{Q}$, while the angle parameters $\{ \zeta_k^{e}, \zeta_k^a \}_{k=0}^K$ are only involved in the former as expressed in \eqref{eq-AoD} and \eqref{eq-AoA}. Accordingly, we decouple the 6D information acquisition problem into two subproblems. In the first subproblem, we estimate the IRS location $\mathbf{p}_{\rm I}$ based on $\{ \hat{\zeta}_k^{e}, \hat{\zeta}_k^a \}_{k=0}^K$; while in the second subproblem, its orientation information $\mathbf{Q}$ is estimated based on both the estimated IRS's location and the angle parameters $\{ \hat{\eta}_k^a, \hat{\eta}_k^e \}_{k=1}^{K}$.

\vspace{-2.5pt}
\subsection{Location Estimation}
First, to estimate the IRS location $\mathbf{p}_{\rm I}$, note that the effective spatial frequencies $\{ \zeta_k^e, \zeta_k^a \}_{k=0}^K  $ in \eqref{zeta0} and \eqref{zetak} can be rewritten in terms of $\mathbf{p}_{\rm I}$ as
\begin{align}
    \zeta_0^e(\mathbf{p}_{\rm I})     =& \frac{2d_t}{\lambda}\frac{(\mathbf{p}_{\rm I} - \mathbf{p}_{\rm TX} )^T \boldsymbol{e}_3}{\| \mathbf{p}_{\rm I} - \mathbf{p}_{\rm TX} \|_2 } , \;\;
{\zeta}_k^e (\mathbf{p}_{\rm I})  =\frac{2d_r}{\lambda} \frac{(\mathbf{p}_{\rm I} - \mathbf{p}_{{\rm RX},k})^T \boldsymbol{e}_3}{\| \mathbf{p}_{\rm I} - \mathbf{p}_{{\rm RX},k}\|_2 } , \forall k=1, \ldots, K   \\
  {\zeta}_0^a(\mathbf{p}_{\rm I})  =& \frac{2d_t}{\lambda} \frac{(\mathbf{p}_{\rm I} - \mathbf{p}_{\rm TX} )^T \boldsymbol{e}_2}{\| \mathbf{p}_{\rm I} - \mathbf{p}_{\rm TX} \|_2 },
\;\;
{\zeta}_k^a(\mathbf{p}_{\rm I})  =  \frac{2d_r}{\lambda}\frac{(\mathbf{p}_{\rm I} - \mathbf{p}_{{\rm Rx},k})^T \boldsymbol{e}_2}{\| \mathbf{p}_{\rm I} - \mathbf{p}_{{\rm Rx},k}\|_2 }, \forall k=1,\ldots, K \label{ang-zetak}
\end{align}
where $\boldsymbol{e}_i \in \mathbb R^{3 \times 1}$ denotes the $i$-th column of the identity matrix $\boldsymbol{I}_3$.

By stacking the actual and estimated angle parameters at the TX and all RXs, we define $\boldsymbol{\zeta}^e(\mathbf{p}_{\rm I}) \triangleq [\zeta_0^e(\mathbf{p}_{\rm I}),\zeta_1^e(\mathbf{p}_{\rm I}), \ldots, \zeta_K^e(\mathbf{p}_{\rm I})]^T $, $\boldsymbol{\zeta}^a(\mathbf{p}_{\rm I}) \triangleq [\zeta_0^a(\mathbf{p}_{\rm I}), \zeta_1^a(\mathbf{p}_{\rm I}), \ldots, \zeta_K^a(\mathbf{p}_{\rm I})]^T $, $\boldsymbol{\hat \zeta}^e \triangleq [\hat{\zeta}_0^e,\hat{ \zeta}_1^e, \ldots, \hat{\zeta}_K^e]^T$, and $\boldsymbol{\hat{\zeta}}^a \triangleq [\hat{\zeta}_0^a, \hat{\zeta}_1^a, \ldots, \hat{\zeta}_K^a]^T.$ Then, the IRS's location $\mathbf{p}_{\rm I}$ can be estimated based on the following least-square estimator,
\begin{align}
\mathbf{\hat p}_{\rm I} = \arg \min_{\mathbf{p}} \quad  & \| \boldsymbol{\zeta}^e(\mathbf{p} ) -  \boldsymbol{\hat{\zeta}}^e \|_2^2  + \| \boldsymbol{\zeta}^a(\mathbf{p} ) -  \boldsymbol{\hat{\zeta}}^a \|_2^2, \quad
 {\text{s.t.}} \quad  \mathbf{p } \in \mathcal{S},
 \label{Opt-Loc}
\end{align}
where $\mathcal{S}$ denotes the set of all possible locations of the IRS.
\vspace{-1pt}

Problem \eqref{Opt-Loc} can be solved by invoking the gradient descent method. In particular, we first ignore the constraint in \eqref{Opt-Loc} and apply the gradient descent method to obtain a converged solution, which is then projected to the feasible set of problem \eqref{Opt-Loc}, i.e., $\cal S$. To accelerate the convergence, we can adopt the Taylor-series expansion to successively approximate its objective function \cite{Foy76}.
Note that for any given local point $\mathbf{\hat p}$, we have
\begin{align}
\zeta_k^e(\mathbf{p}) - \hat{\zeta}_k^e =& \zeta_k^e( \hat{\mathbf{p}}) - \hat{\zeta}_k^e   + ( \mathbf{p} - \hat{\mathbf{p}})^T \frac{\partial \zeta_k^e(\mathbf{p})}{\partial \mathbf{p}}\bigg|_{\mathbf{p} = \hat{\mathbf{p}}}  + {o}( \| \mathbf{p} - \hat{\mathbf{p}} \|_2^2),
\label{eq-Tayl}
\end{align}
where $o(\cdot)$ denotes the higher-order terms, which would vanish when $\mathbf{\hat p}$ approaches $\mathbf{p}$, and the first-order derivative of $\zeta_k^e(\mathbf{p}_{})$ w.r.t. $\mathbf{p}$ in \eqref{eq-Tayl} is calculated as 
\begin{align}
\frac{\partial \zeta_k^e(\mathbf{p})}{\partial \mathbf{p}} =  \begin{cases}
 \frac{2d_t}{\lambda} {\boldsymbol{f} }(\mathbf{p}; \boldsymbol{e}_3, \mathbf{p}_{\rm TX}) , & k=0 \\
\frac{2d_r}{\lambda} {\boldsymbol{f} }(\mathbf{p}; \boldsymbol{e}_3, \mathbf{p}_{{\rm RX},k}) , & k=1,2,\ldots, K ,
\end{cases} 
\label{eq-deri}
\end{align}
where ${\boldsymbol{f} }(\mathbf{p}; \boldsymbol{e}, \mathbf{p}_k) \triangleq \frac{  \| \mathbf{p}- \mathbf{p}_k \|_2^2 \boldsymbol{e} - (\mathbf{p}- \mathbf{p}_k)^T \boldsymbol{e} (  \mathbf{p}-\mathbf{p}_k) }{  \| \mathbf{p} - \mathbf{p}_k \|_2^3}$ is a vector function in terms of $\mathbf{p}$. The same procedure as in \eqref{eq-Tayl} can be performed over $\zeta_k^a(\mathbf{p}$), where its first-order derivative $\frac{\partial \zeta_k^a (\mathbf{p})}{\partial \mathbf{p}}$ can be obtained by replacing $\boldsymbol{e}_3$ in \eqref{eq-deri} with $\boldsymbol{e}_2$. 
For convenience, by stacking the first-order derivatives of $\{\zeta_k^e(\mathbf{p}), \zeta_k^a(\mathbf{p}) \}_{k=0}^K$ w.r.t. $\mathbf{p}$, we define
\begin{align}
\boldsymbol{A} (\mathbf{\hat p}) \triangleq &\bigg[
 \frac{\partial \zeta_0^e(\mathbf{p})}{\partial \mathbf{p}}\bigg|_{\mathbf{p} = \hat{\mathbf{p}}} , \ldots, \frac{\partial \zeta_K^e(\mathbf{p})}{\partial \mathbf{p}}\bigg|_{\mathbf{p} = \hat{\mathbf{p}}},  \frac{ \partial \zeta_0^a(\mathbf{p})}{\partial \mathbf{p}}\bigg|_{\mathbf{p} = \hat{\mathbf{p}}}, \ldots, \frac{ \partial \zeta_K^a(\mathbf{p})}{\partial \mathbf{p}}\bigg|_{\mathbf{p} = \hat{\mathbf{p}}}
 \bigg] \in \mathbb{R}^{3 \times 2(K+1)} .
 \label{eq-A}
\end{align}
By substituting \eqref{eq-Tayl}-\eqref{eq-A} into the objective function of \eqref{Opt-Loc} and ignoring the higher-order terms, it can be approximated as
\begin{align}
 &\| \boldsymbol{\zeta}^e(\mathbf{p} ) -  \boldsymbol{\hat{\zeta}}^e \|_2^2  + \| \boldsymbol{\zeta}^a(\mathbf{p} ) -  \boldsymbol{\hat{\zeta}}^a \|_2^2  
 \approx   \| \boldsymbol{\Delta}_g (\mathbf{\hat p})- \boldsymbol{A}^T (\mathbf{\hat p}) (\mathbf{p}-\mathbf{\hat p} )\|_2^2,
 \label{eq-Obj}
\end{align}where $\boldsymbol{\Delta}_g (\mathbf{\hat p}) \triangleq [ (\boldsymbol{\hat \zeta}^e)^T , (\boldsymbol{\hat \zeta}^a)^T ]^T - [(\boldsymbol{\zeta}^e(\hat {\mathbf{p}}))^T, (\boldsymbol{\zeta}^a(\hat {\mathbf{p}} ))^T ]^T $.
 
Let $\mathbf{p}^{(t)}$ denote the optimized $\mathbf{p}$ of the gradient descent method in its $t$-th iteration. Then, the iteration can proceed as
\begin{align}
\mathbf{p}^{(t+1)} = \arg \min_{\mathbf{p}}  \| \boldsymbol{\Delta}_g (\mathbf{p}^{(t)})- \boldsymbol{A}^T (\mathbf{p}^{(t)})  (\mathbf{p}-\mathbf{p}^{(t)})\|_2^2,
\end{align}
which admits a closed-form solution given by
\begin{align}
\mathbf{ p}^{(t+1)}= \mathbf{p}^{(t)} +  ( \boldsymbol{A} (\mathbf{p}^{(t)})\boldsymbol{A}^T(\mathbf{p}^{(t)}) )^{-1} \boldsymbol{A} (\mathbf{p}^{(t)})\boldsymbol{\Delta}_g (\mathbf{p}^{(t)}).
\end{align}
The gradient descent method can be iteratively performed until the convergence condition $\|\mathbf{p}-\mathbf{p}^{(t)} \| < \varepsilon_0 $ is met, where $\varepsilon_0$ is a preset stopping threshold. Finally, we project the converged solution onto the set $\mathcal{S}$ as the estimate of the IRS's location.





\subsection{Orientation Estimation}
\label{sec-ori}
Let $\mathbf{\hat p}_{\rm I}$ denote the estimate of the IRS's location by the proposed algorithm. With $\mathbf{\hat p}_{\rm I}$ and the estimated cascaded spatial frequencies at the IRS, i.e., $ \boldsymbol{\hat \eta}^e \triangleq [\hat {\eta}_1^e, \hat {\eta}_2^e, \ldots, \hat {\eta}_K^e]^T$ and $\boldsymbol{\hat {\eta}}^a \triangleq [\hat {\eta}_1^a, \hat {\eta}_2^a, \ldots, \hat {\eta}_K^a]^T$, we next show how to estimate the rotation matrix $\mathbf{Q}$.

Specifically, we first rewrite the cascaded azimuth/elevation spatial frequencies at the IRS as
\begin{align}
\eta^a_k  = &\frac{2  d_{I}}{\lambda}  \bigg( \frac{q_{A,x} } { \| \mathbf{q}_A \|_2} 
- \frac{q_{D,x} } { \| \mathbf{q}_D \|_2}  \bigg) = \boldsymbol{e}_1^T \mathbf{Q}^T \boldsymbol{b}_k(\mathbf{p}_{\rm I})  ,\forall k,   \label{eq-etaqa}\\
\eta^e_k   = & \frac{2  d_{I}}{\lambda}  \bigg( \frac{q_{{A},y} } { \| \mathbf{q}_A \|_2}   
- \frac{q_{D_k,y} } { \| \mathbf{q}_{D_k} \|_2} \bigg) = \boldsymbol{e}_2^T \mathbf{Q}^T \boldsymbol{b}_k (\mathbf{p}_{\rm I}) , \forall k,  \label{eq-etaqe}
\end{align}
where
\begin{align}
\boldsymbol{b}_k(\mathbf{p}_{\rm I})  =& \frac{2  d_{I}}{\lambda}  \left( \frac{\mathbf{p}_{\rm I} -\mathbf{p}_{\rm TX} }{\| \mathbf{p}_{\rm I} -\mathbf{p}_{\rm TX}\|_2} -  \frac{\mathbf{p}_{{\rm RX},k}- \mathbf{p}_{\rm I} }{\| \mathbf{p}_{{\rm RX},k}- \mathbf{p}_{\rm I}\|_2} \right).
\label{eq-bk}
\end{align}
Let $ \boldsymbol{B}(\mathbf{p}_{\rm I}) \triangleq [ \boldsymbol{b}_1(\mathbf{p}_{\rm I}), \cdots, \boldsymbol{b}_K(\mathbf{p}_{\rm I})] \in \mathbb R^{3 \times K}$, $\boldsymbol{\eta}^a = [ \eta_1^a, \ldots, \eta_K^a]^T \in \mathbb R^{ K \times 1}$ and $\boldsymbol{\eta}^e = [ \eta_1^e, \ldots, \eta_K^e]^T \in \mathbb R^{ K \times 1}$. Then, \eqref{eq-etaqa} and \eqref{eq-etaqe} can be rewritten into a more compact form,
\begin{align}
\boldsymbol{B}^T(\mathbf{p}_{\rm I}) \mathbf{Q} \boldsymbol{e}_1 = \boldsymbol{\eta}^a,   \quad \boldsymbol{B}^T(\mathbf{p}_{\rm I}) \mathbf{Q} \boldsymbol{e}_2 = \boldsymbol{\eta}^e ,
\label{eq-B}
\end{align}
respectively, and stacking them yields
\begin{align}
\boldsymbol{B}^T(\mathbf{p}_{\rm I})\mathbf{Q} \boldsymbol{E} = \boldsymbol{T},
\label{eq-BQ}
\end{align}
where $\boldsymbol{E} \triangleq [ \boldsymbol{e}_1, \boldsymbol{e}_2] \in \mathbb R^{3 \times 2}$ and $\boldsymbol{T} \triangleq [ \boldsymbol{\eta}^a, \boldsymbol{\eta}^e] \in \mathbb R^{3 \times 2}$.

Next, let $\boldsymbol{\hat B} =\boldsymbol{B}(\hat{\mathbf{p}_{\rm I}})$ and $\boldsymbol{\hat T} =  [ \boldsymbol{\hat{\eta}}^a, \boldsymbol{\hat{\eta}}^e]$ denote the reconstructed $\boldsymbol{B}$ and $\boldsymbol{T}$ with the estimated IRS's location $\mathbf{\hat p}_{\rm I}$ and cascaded spatial frequencies $\{ \hat{\eta}_k^a, \hat{\eta}_k^e \}_{k=1}^K$, respectively. Based on \eqref{eq-BQ}, we can estimate $\mathbf{Q}$ by solving the following least-square problem,
\begin{align}
\hat {\mathbf{Q}} = \quad &\arg \min_{\mathbf{Q}} \| \boldsymbol{ \hat B}^T \mathbf{Q} \boldsymbol{E} - \boldsymbol{\hat T} \|_F^2 \nonumber \\
 {\text{s.t.}} \quad  & \mathbf{Q} \in {\text{SO}}(3) =  \{\mathbf{Q}:{\det}(\mathbf{Q})=1, \mathbf{Q}\mathbf{Q}^T = \boldsymbol{I}_3 \}.
\label{opt-2}
\end{align}
It is worth mentioning that as long as $K\geq 2$, there would exist at least four independent cascaded elevation and azimuth spatial frequencies, which is sufficient to ensure that the solution to \eqref{opt-2} is unique, as the number of unknown variables in $\mathbf{Q}$ is only equal to three. 
However, problem \eqref{opt-2} is a non-convex optimization problem due to the 3D rotation constraints involved. In the following, we first propose an iterative manifold optimization algorithm \cite{Boumal2020} to solve \eqref{opt-2} locally. Then, in the special case that $\boldsymbol{\hat B}$ has a full rank, we show that problem \eqref{opt-2} admits a closed-form solution.

Specifically, in the manifold optimization, we first rewrite the objective function in \eqref{opt-2} as
\begin{align}
{g}(\mathbf{Q}) \triangleq  & \| \boldsymbol{ \hat B}^T \mathbf{Q} \boldsymbol{E} - \boldsymbol{ \hat T} \|_F^2   =  {\rm Tr} \big( \mathbf{Q}^T \boldsymbol{ \hat B} \boldsymbol{ \hat B}^T \mathbf{Q} \boldsymbol{E}\boldsymbol{E}^T  \big) -2 {\rm Tr} \big(  \mathbf{Q}^T \boldsymbol{ \hat B}  \boldsymbol{ \hat T}  \boldsymbol{E}^T \big)  + {\rm Tr}(\boldsymbol{ \hat T} \boldsymbol{ \hat T}^T).
\end{align}
Let $\mathbf{Q}^{(s)}$ denote the optimized $\mathbf{Q}$ in the $s$-th iteration of the manifold optimization. Then, the iteration can proceed as \cite{Boumal2020}
\begin{align}
\mathbf{ Q}^{(s+1)} = {\text{Ret}}_{ \mathbf{Q}^{(s)}} \bigg( -\upsilon_s {\text{Proj}}_{\mathbf{Q}^{(s)}}  \frac{\partial {g}( \mathbf{Q})}{ \partial \mathbf{Q}}\bigg),
\label{eq-updQ}
\end{align}
where ${\text{Proj}}_{\mathbf{Q}^{(s)}} (\cdot)$ is the projection of its argument onto the tangent space of $\mathbf{Q}^{(s)}$, $\upsilon_s >0$ is the step size in the $s$-th iteration, and ${\rm Ret}_{\mathbf{ Q}^{(s)}}(\cdot)$ is a retraction from the tangent space of $\mathbf{Q}^{(s)}$ onto $\text{SO}(3)$.
The above components of \eqref{eq-updQ} can be computed as follows. The Euclidean gradient $\frac{\partial {g}(\mathbf{Q})}{ \partial \mathbf{Q}}$ is given by
\begin{align}
\frac{\partial {g} (\mathbf{Q})}{\partial \mathbf{Q}} = 2\boldsymbol{ \hat B}\boldsymbol{ \hat B}^T \mathbf{Q} \boldsymbol{E}\boldsymbol{E}^T - 2 
\boldsymbol{ \hat B}  \boldsymbol{ \hat T}  \boldsymbol{E}^T,
\end{align}the projection operator is given by 
\begin{align}
{\rm Proj}_{\boldsymbol{X}} (\boldsymbol{U}) = \boldsymbol{X} ({\boldsymbol{X}^T \boldsymbol{U} - \boldsymbol{U}^T \boldsymbol{X} })/{2}     ,
\end{align}
and the retraction operator is given by
\begin{align}
{\rm Ret}_{\boldsymbol{X}}(\boldsymbol{U}) = (\boldsymbol{X} + \boldsymbol{U})(\boldsymbol{I}_3 + \boldsymbol{U}^T \boldsymbol{U})^{ -1/2}.
\end{align}As shown in \cite{Boumal2020}, the iteration in \eqref{eq-updQ} is guaranteed to converge to a critical point of \eqref{opt-2}.

Furthermore, note that in the special case that $\boldsymbol{\hat B}$ has a full row rank, problem \eqref{opt-2} can be equivalently recast as
\begin{align}
\hat {\mathbf{Q}} = \quad &\arg \min_{\mathbf{Q}} \| \mathbf{Q} \boldsymbol{E} -  (\boldsymbol{ \hat B}^T )^{\dagger}  \boldsymbol{ \hat T} \|_F^2, \quad 
{\text{s.t.}} \quad  \mathbf{Q} \in {\text{SO}}(3) ,
\label{opt-3}
\end{align}
where $(\boldsymbol{\hat B}^T)^{\dagger} =  ( \boldsymbol{\hat B} \boldsymbol{\hat B} ^T)^{-1} \boldsymbol{\hat B} $ is the Moore-Penrose pseudo-inverse of the matrix $\boldsymbol{\hat B}^T $.

Problem \eqref{opt-3} can be optimally solved in closed-form by invoking the Kabsch algorithm \cite{Kabsch76}.
In particular, we conduct singular value decomposition (SVD) on the matrix $ \boldsymbol{A} \triangleq  \boldsymbol{E} \boldsymbol{ \hat T}^T \boldsymbol{ \hat B} ^{\dagger}  \in \mathbb R^{3 \times 3} $ as $\boldsymbol{A} = \boldsymbol{U} \boldsymbol{\Sigma} \boldsymbol{V}^T$.
Then, the optimal solution to \eqref{opt-3} can be obtained as $\mathbf{\hat Q} = {\rm det}(\boldsymbol{V}\boldsymbol{U}^T)\boldsymbol{V} \boldsymbol{U}^T$ \cite{Kabsch76},
which satisfy the 3D rotation constraint in $\text{SO}(3)$.

It is also worth noting that by properly determining the sensing RXs’ locations, if $K \geq 3$, the matrix $\boldsymbol{\hat B}$ should have a full row rank with a high probability. Therefore, the above special case can usually hold in practice when $K \geq 3$, which helps improve the accuracy of IRS orientation estimation and also makes the computation easier.

\section{Theoretical Analysis}
\label{sec-sens}
\allowbreak
In this section, we first analyze the performance of the proposed sensing scheme by deriving the CRBs on the estimates of the involved angle parameters $\{ \zeta_0^a,\zeta_0^e, \zeta_k^a, \zeta_k^e,\eta_k^a, \eta_k^e \}$, the IRS's location $\mathbf{ p}_{\rm I}$, as well as the rotation matrix $\mathbf{Q}$. Furthermore, to derive useful insights, we analyze the sensitivity of the orientation estimation to the estimation errors in the IRS's location and cascaded angle parameters.

\subsection{CRBs in 6D Information Acquisition}
\label{Section-CRB}

First, we group the estimates of all associated angle parameters for 6D information acquisition as
\begin{align}
\boldsymbol{\Gamma} = [(\boldsymbol{  \zeta}^a)^T, (\boldsymbol{\zeta}^e)^T, (\boldsymbol{\eta}^a)^T, (\boldsymbol{ \eta}^e)^T]^T  \in \mathbb R^{(4K+2) \times 1 }.
\label{eq-Gamma}
\end{align}The CRB on the angle parameter vector in \eqref{eq-Gamma} can be derived as the inverse of its Fisher information matrix (FIM), denoted as $\boldsymbol{I}(\boldsymbol{\Gamma})$, for which the details are provided in Appendix \ref{AppendixA}.

With the FIM $\boldsymbol{I}(\boldsymbol{\Gamma})$, we can derive the CRBs on the IRS's location $\mathbf{ p}_{\rm I}$ and orientation $\mathbf{Q}$ by exploiting the chain rule \cite{Kay93}. First, as shown in \eqref{eq-Qseq}, the rotation matrix $\mathbf{Q}$ is dependent on the axis angle vector $\boldsymbol{\psi}$. Hence, we introduce an intermediate vector $\mathbf{s} \triangleq [ \mathbf{p}_{\rm I} ^T, \boldsymbol{\psi} ^T] ^T\in \mathbb R^{6 \times 1}$, and the FIM of the intermediate vector $\mathbf{s}$ can be obtained based on the chain rule as 
\begin{align}
\boldsymbol{I}(\mathbf{s}) = (\nabla_{\boldsymbol{s}} \boldsymbol{\Gamma})^T \boldsymbol{I}(\boldsymbol{\Gamma}) (\nabla_{\boldsymbol{s}} \boldsymbol{\Gamma}),
\end{align}where $ \nabla_{\boldsymbol{s}} \boldsymbol{\Gamma}  \in \mathbb R^{ (4K+2) \times 6}$ is the Jacobian matrix of the vector $\boldsymbol{\Gamma}$ w.r.t. $\mathbf{s}$, whose expression is given in Appendix \ref{AppendixB}.

Therefore, the CRB on the IRS's location $\mathbf{p}_{\rm I}$ can be written as
\begin{align}
{\text{CRB}} (\mathbf{p}_{\rm I})= {\rm tr} \left(     \boldsymbol{I}^{-1}(\mathbf{s})_{1:3,1:3} \right),
\label{eq-PEB}
\end{align}where $[\cdot]_{a:b,c:d}$ indicates the sub-matrix of the argument located between rows $(a,b)$ and columns $(c,d)$. Second, for the rotation matrix $\mathbf{Q}$, its CRB can be derived based on its vectorized version, $\mathbf{\bar q} \triangleq {\rm vec}(\mathbf{Q})$, which is given by \cite{Kay93}
\begin{align}
{\text {CRB} }(\mathbf{Q}) = {\rm tr}\bigg( (\nabla_{\boldsymbol{s}} \mathbf{\bar q})  \boldsymbol{I}^{-1}(\mathbf{s} ) (\nabla_{\boldsymbol{s}} \mathbf{\bar q})^T\bigg),  
\label{eq-OEB}
\end{align}
where $ \nabla_{\boldsymbol{s}} \mathbf{\bar q} \in \mathbb R^{ 9 \times 6}$ is the Jacobian matrix of the vector $\mathbf{\bar q}$ w.r.t. $\mathbf{s}$, whose expression is given in Appendix \ref{AppendixC}.

\subsection{Sensitivity Analysis of Orientation Estimation}
\label{sec-sense2}

Although the CRBs derived in Section \ref{Section-CRB} capture the minimum variance of the estimation error by any unbiased estimator, they cannot explicitly quantify the impact of the angle parameters/location estimation on the subsequent orientation estimation, which is pursued in this subsection. In particular, we aim to analyze how the localization error, $\mathbf{\Delta} \mathbf{p}\triangleq \mathbf{\hat p}_I -  \mathbf{ p}_I$, and the estimation errors of cascaded spatial frequencies in \eqref{eq-BQ}, $\boldsymbol{\Delta}\boldsymbol{T}\triangleq \boldsymbol{\hat T} - \boldsymbol{T}$, affect the accuracy of orientation estimation.

To this end, we first analyze the effect of localization error $\boldsymbol{\Delta }{\mathbf{p}}$ on an intermediate parameter, i.e., the matrix $\boldsymbol{B}$ in \eqref{eq-B}. 
In particular, given the estimated IRS's location $\mathbf{\hat p}_{\rm I}$, $\boldsymbol{B}$ can be reconstructed as $\boldsymbol{\hat B}=[\boldsymbol{\hat b}_1, \boldsymbol{\hat b}_2, \cdots, \boldsymbol{\hat b}_{K}]$, with
\begin{align}
\mathbf{\hat b}_k  \triangleq & \frac{\mathbf{ \hat p}_{\rm I} -\mathbf{p}_{\rm TX} }{\| \mathbf{\hat p}_{\rm I} -\mathbf{p}_{\rm TX}\|_2} -  \frac{\mathbf{p}_{{\rm RX},k}- \mathbf{\hat p}_{\rm I} }{\| \mathbf{p}_{{\rm RX},k}- \mathbf{\hat p}_{\rm I}\|_2}  =  \frac{\mathbf{ p}_{\rm I} + \boldsymbol{\Delta} \mathbf{p} -\mathbf{p}_{\rm TX} }{\| \mathbf{ p}_{\rm I} + \boldsymbol{\Delta} \mathbf{p} -\mathbf{p}_{\rm TX}\|_2} - \frac{\mathbf{p}_{{\rm RX},k}- \mathbf{ p}_{\rm I} - \boldsymbol{\Delta }\mathbf{p} }{\| \mathbf{p}_{{\rm RX},k}- \mathbf{p}_{\rm I} - \boldsymbol{\Delta }\mathbf{p}  \|_2} .
\end{align}
Furthermore, note that the TX-IRS and IRS-RX distances are generally much larger than the localization error (usually at centimeter levels \cite{WangZhang21Joint}), which results in $\| \mathbf{ p}_{\rm I} + \boldsymbol{\Delta} \mathbf{p} -\mathbf{p}_{\rm TX}\|_2 \approx \| \mathbf{ p}_{\rm I}   -\mathbf{p}_{\rm TX}\|_2$, and $\| \mathbf{p}_{{\rm RX},k}- \mathbf{p}_{\rm I} - \boldsymbol{\Delta }\mathbf{p}  \|_2\approx \| \mathbf{p}_{{\rm RX},k}- \mathbf{p}_{\rm I}  \|_2 $. As such, we have
\begin{align}
\mathbf{\hat b}_k  
\approx  
& \boldsymbol{b}_k + \bigg( \frac{ 1 }{\| \mathbf{\hat p}_{\rm I} -\mathbf{p}_{\rm TX}\|_2} +  \frac{1 }{\| \mathbf{p}_{{\rm RX},k}- \mathbf{\hat p}_{\rm I}\|_2}  \bigg) \boldsymbol{\Delta} \mathbf{p} = \boldsymbol{b}_k + r _k\boldsymbol{\Delta}\mathbf{p} ,
\label{eq-rk}
\end{align}
where $r_k \triangleq  \frac{ 1 }{\| \mathbf{\hat p}_{\rm I} -\mathbf{p}_{\rm TX}\|_2} +  \frac{1 }{\| \mathbf{p}_{{\rm RX},k}- \mathbf{\hat p}_{\rm I}\|_2} \ll 1$.
By defining $\boldsymbol{r} =[r_1, \ldots, r_K]^T$, the estimation error of $\boldsymbol{B}$ can be expressed in terms of $\boldsymbol{\Delta} \mathbf{p}$ as $\boldsymbol{\Delta} \boldsymbol{B} = \boldsymbol{\hat B} -\boldsymbol{B} \approx  \boldsymbol{  \Delta} \mathbf{p}  \times \boldsymbol{r}^T$,
which yields
\begin{align}
\| \boldsymbol{\Delta} \boldsymbol{B} \|_2 \approx \| \boldsymbol{\Delta } \mathbf{p}\|_2 \|\boldsymbol{r}\|_2.
\label{eq-dB2}
\end{align}
Next, we analyze how $\boldsymbol{\Delta B}$ affects the orientation estimation in solving \eqref{opt-2}. To simplify the analysis, we consider that the matrix $\boldsymbol{B}$ has a full row rank (corresponding to the special case presented in Section \ref{sec-ori}), and the IRS's location and cascaded spatial frequencies are perfectly estimated. As a result, the rotation matrix $\mathbf{Q}$ satisfies
\begin{align}
 \mathbf{Q} \boldsymbol{E} =  (\boldsymbol{  B}^T )^{\dagger}  \boldsymbol{  T}, 
\quad & \mathbf{Q} \in {\text{SO}}(3) .
\label{opt-Q4}
\end{align}As $\boldsymbol{E}=[\boldsymbol{e}_1,\boldsymbol{e}_2]$, we can further obtain
\begin{align}
\mathbf{q}_1 =   (\boldsymbol{  B}^T )^{\dagger} \boldsymbol{t}_1, 
\mathbf{q}_2 =   (\boldsymbol{  B}^T )^{\dagger} \boldsymbol{t}_2, 
\label{eq-q12}
\end{align}where $\mathbf{q}_i$ and $\boldsymbol{t}_{i},i=1,2$ denote the $i$-th column of the matrix $\mathbf{Q}$  and $\boldsymbol{T}$, respectively.
Although the third column of $\mathbf{Q}$, i.e., $\mathbf{q}_3$, cannot be obtained directly by this means, it can be inferred that $\mathbf{q}_3$ can be expressed as a similar form to \eqref{eq-q12}. This can be explained by assuming a virtual IRS lying in the $y'$-$z'$ plane of the local CCS in Fig. \ref{System_fig} with the same location as the actual one. Then, by applying the proposed orientation estimation method to this virtual IRS, we can still obtain \eqref{opt-Q4}, whereas $\boldsymbol{E}=[\boldsymbol{e}_1,\boldsymbol{e}_2]$ therein should be replaced with $\boldsymbol{E}=[\boldsymbol{e}_2,\boldsymbol{e}_3]$, and this can yield a similar form to \eqref{eq-q12} for $\mathbf{q}_{\rm 3}$. Hence, any column in $\mathbf{Q}$ can be represented as a general form of $ \mathbf{q} = (\boldsymbol{B}^T)^{\dagger} \boldsymbol{t}$,
where $\boldsymbol{t} \in \mathbb R^{K \times 1}$ is the associated cascaded spatial frequencies.
To further simplify the analysis and gain more insights, we assume $K=3$, which results in $ (\boldsymbol{B}^T)^{\dagger} = (\boldsymbol{  B}^T )^{-1} $ and
\begin{align}
\mathbf{q} =  (\boldsymbol{  B}^T )^{-1} \boldsymbol{t}.
\label{eq-q}
\end{align}
Based on \eqref{eq-q}, we present the following lemma \cite{Golub13Matrix}.
\begin{lemma} 
Let $\boldsymbol{\Delta B}$ and $\boldsymbol{\Delta t}$ denote the perturbation on the matrix $\boldsymbol{B}$ and the vector $\boldsymbol{t}$, respectively. Then, the estimate of $\mathbf{q}$ with the above perturbations is given by
\begin{align}
\mathbf{\hat q}  =( (\boldsymbol{  B} + \boldsymbol{\Delta} \boldsymbol{  B} )^T )^{-1} (\boldsymbol{t} + \boldsymbol{\Delta}  \boldsymbol{t} ).
\end{align}
If $\| (\boldsymbol{B}^T)^{-1}\|_2 \| \boldsymbol{ \Delta B} \|_2 < 1$, the relative error between $\mathbf{\hat q}$ and $\mathbf{q}$ is upper-bounded by
\begin{align}
\frac{ \| \hat {\mathbf{q}} - \mathbf{q} \|_2 }{ \| \mathbf{q} \|_2} \leq &\frac{\kappa(\boldsymbol{B})}{1- \kappa(\boldsymbol{B}) \| \boldsymbol{\Delta B} \|_2/ \| \boldsymbol{B} \|_2 }  \left( \frac{ \| \Delta \boldsymbol{B}\|_2 }{ \| \boldsymbol{B} \|_2}   +  \frac{ \| \boldsymbol{\Delta } \boldsymbol{t} \| _2}{  \| \boldsymbol{t} \|_2} \right) 
\label{eq-bound}
\end{align}
where $\kappa(\boldsymbol{B}) = \|\boldsymbol{B}\|_2 \| \boldsymbol{B}^{-1} \|_2$ denotes the condition number of the matrix $\boldsymbol{B}$. 
\label{theorem-1}
\end{lemma}

Note that thanks to \eqref{eq-dB2}, the prerequisite for \eqref{eq-bound}, $\| (\boldsymbol{B}^T)^{-1}\|_2 \| \boldsymbol{ \Delta B} \|_2 < 1$, in Lemma \ref{theorem-1} can usually be met in practice. It is observed from \eqref{eq-bound} that the upper bound on the relative orientation estimation error critically depends on the condition number $\kappa(\boldsymbol{B})$. When the condition number $\kappa(\boldsymbol{B})$ is large, even a small $\| \boldsymbol{\Delta} \boldsymbol{B} \|_2$ (or $\|\boldsymbol{\Delta } \mathbf{p} \|_2$ based on \eqref{eq-dB2}) can drastically change the estimated rotation matrix. Since the condition number $\kappa(\boldsymbol{B})$ is determined by the locations of the TX, IRS, and all RXs, their deployment should have a significant effect on the orientation estimation, as will be shown in Section \ref{sec-simulation} via simulation results. Furthermore, for any given $\kappa(\boldsymbol{B})$, since $r_{k} \ll 1, \forall k$ in \eqref{eq-dB2}, the right-hand side of \eqref{eq-bound} should be more dominated by its second term, $\|\boldsymbol{\Delta} \boldsymbol{t} \|_2/\| \boldsymbol{t} \|_2 $, than its first term, $\|\boldsymbol{\Delta} \boldsymbol{B}\|_2/\| \boldsymbol{\Delta} \boldsymbol{B}\|_2$. Thus, the orientation estimation may be more sensitive to the estimation error of the cascaded spatial frequencies $\| \boldsymbol{\Delta t} \|_2$ compared to the localization error $\|\boldsymbol{\Delta } \mathbf{p} \|_2$, as will also be numerically verified in Section \ref{sec-simulation} next.




\section{Simulation Results}
\label{sec-simulation}

In this section, we evaluate the performance of our proposed 6D information acquisition methods. Unless otherwise stated, the simulation parameters are set as follows. We consider Rician fading for all channels involved with the Rician factor of $10$ dB. The system carrier frequency is $28$ GHz, and the path-loss exponents of the TX-IRS and IRS-RX $k$ channels are set to $2$ and $2.2$, respectively. The noise power is set to $\sigma^2 = -110$ dBm. The numbers of antennas at the TX, IRS, and RXs are set to $N_{t_y} = 8$, $N_{t_z} = 8$, $M_x=8$, $M_y=8$, $N_{r_y} = 8$, and $N_{r_z} = 8$. Moreover, the antenna/element spacing of the TX, IRS and each RX are set to $d_r =\lambda /2$, $d_I = \lambda/4$, and $d_t = \lambda/2$, respectively. The numbers of beam codewords at the TX, IRS and RXs are $D_{\rm TX} = D_{\rm IRS} = D_{\rm RX}=36$. The location and rotation matrix of the IRS/target are $\mathbf{p}_{\rm I}=[5,4,10]^T$ and $\mathbf{Q} = \mathbf{Q}_z(\frac{\pi}{4}) \mathbf{Q}_y(\frac{\pi}{6}) \mathbf{Q}_x(\frac{\pi}{4})$, respectively. The number of RXs is $K = 2$, and their locations are fixed as $\mathbf{p}_{{\rm RX},1} = [30,-25,9]^T$, $\mathbf{p}_{{\rm RX},2} = [22,27,0]^T$, respectively. The location range of the IRS is set to $\mathcal{S} \triangleq \{ \mathbf{p}=[p_x,p_y,p_z] | |p_{i} -p_{{\rm I},i}| \leq 20, i \in \{ x,y,z\} \}$. All the results to be shown are averaged over $1000$ independent channel realizations. 

\begin{figure}[!t]
 \centering
 \subfigure[MSE and CRB of $\zeta_0^a$.]{
  \begin{minipage}{.31\textwidth}
   \centering
   \includegraphics[scale=.38]{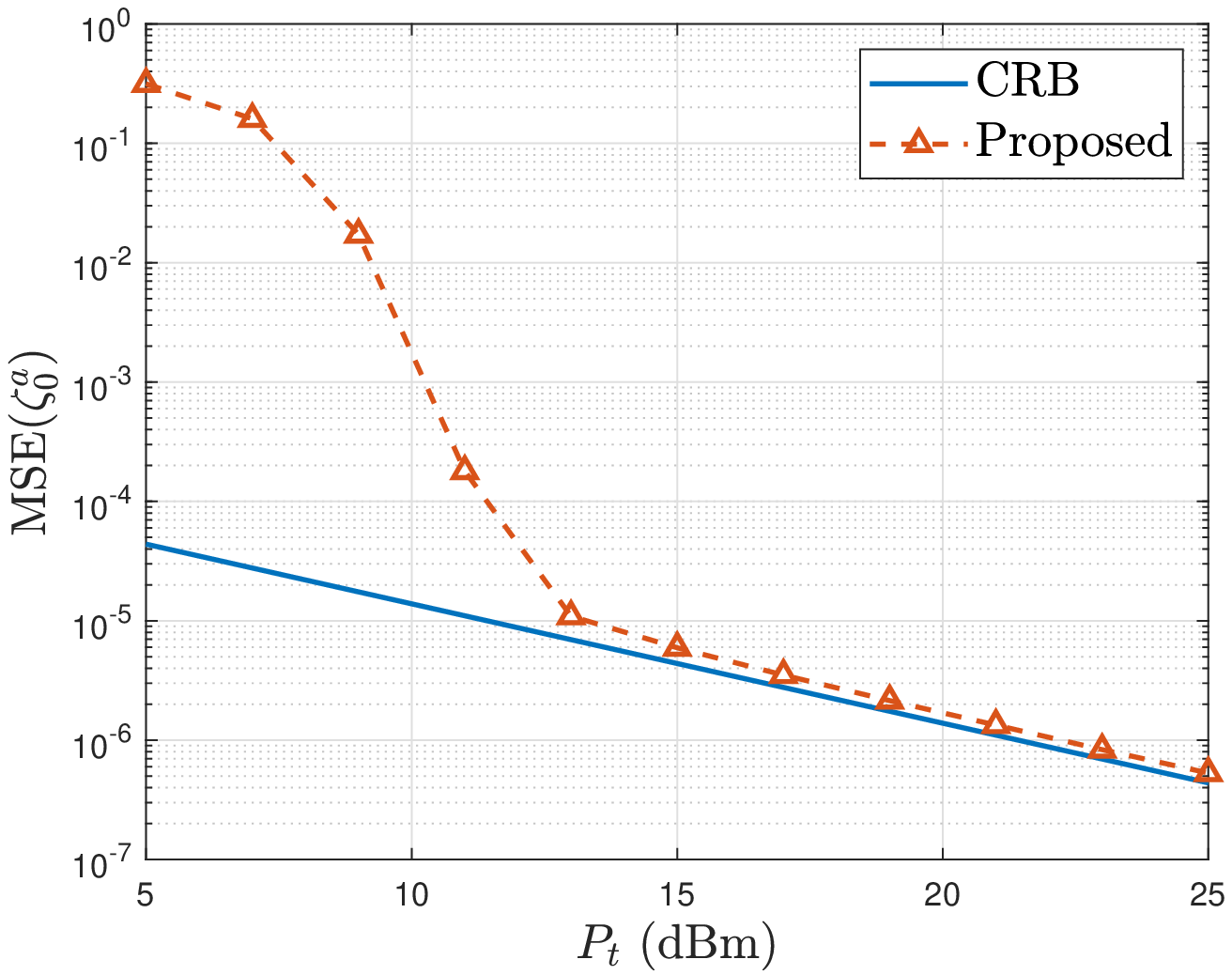}
  \end{minipage}
  \label{FIG6_1}
 }
 \subfigure[MSE and CRB of $\zeta_1^a$.]{
  \begin{minipage}{.31\textwidth}
   \centering
   \includegraphics[scale=.38]{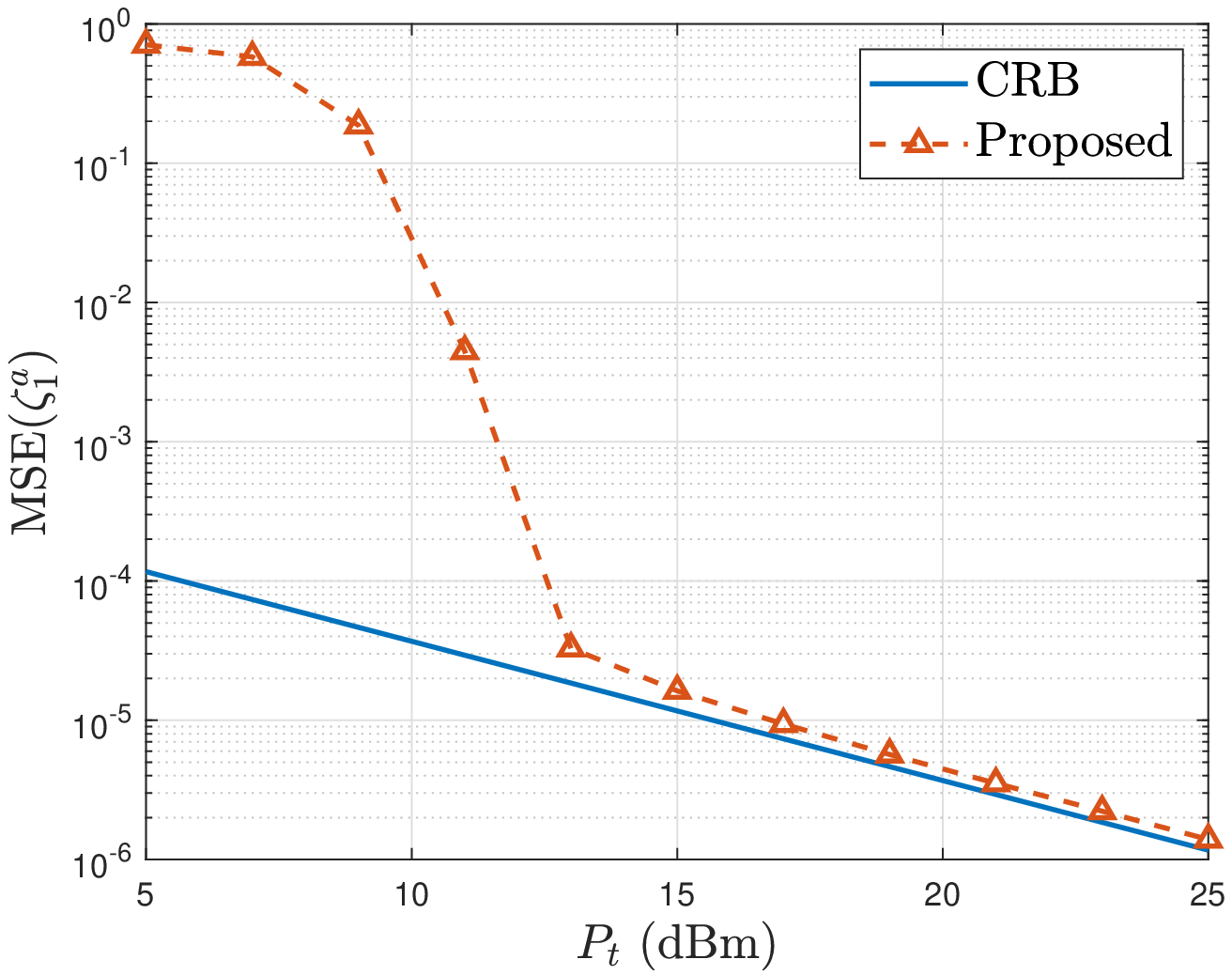}
  \end{minipage}
  \label{FIG6_2}
 }
 \subfigure[MSE and CRB of $\eta_1^a$.]{
  \begin{minipage}{.31\textwidth}
   \centering
   \includegraphics[scale=.38]{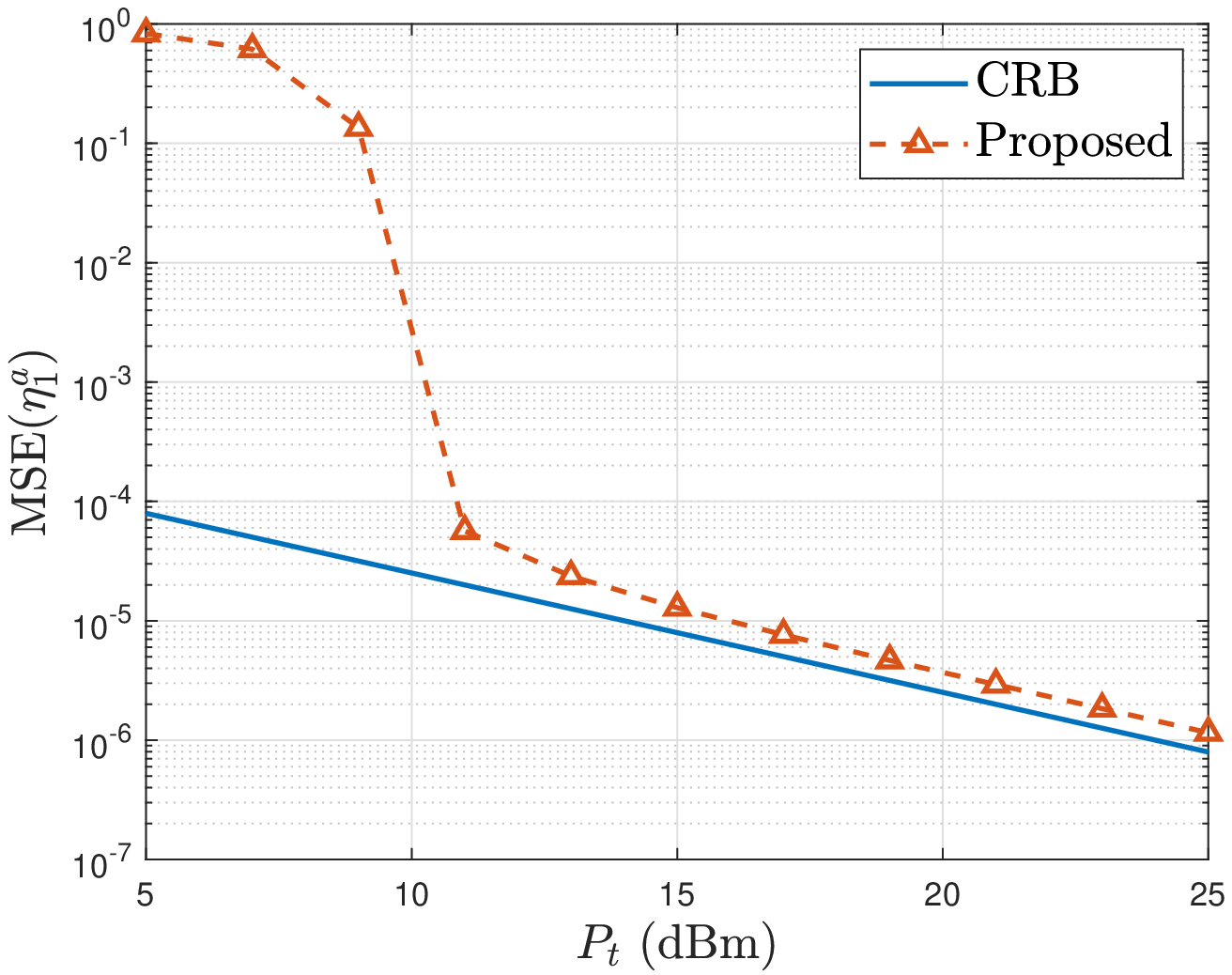}
  \end{minipage}
  \label{FIG6_3}
 }
 \hfill
  \subfigure[MSE and CRB of $\zeta_0^e$.]{
  \begin{minipage}{.31\textwidth}
   \centering
   \includegraphics[scale=.38]{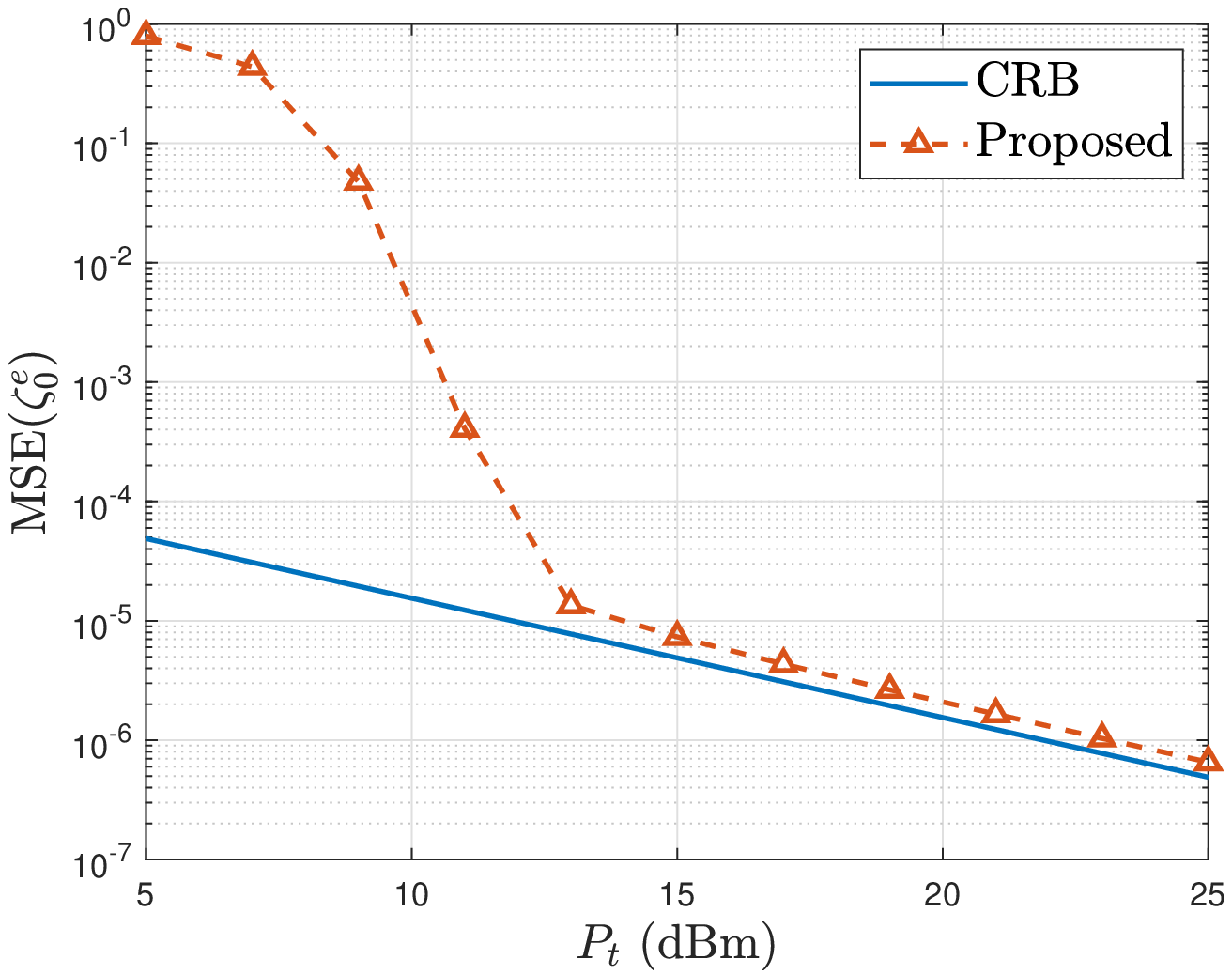}
  \end{minipage}
  \label{FIG6_4}
 }
  \subfigure[MSE and CRB of $\zeta_1^e$.]{
  \begin{minipage}{.31\textwidth}
   \centering
   \includegraphics[scale=.38]{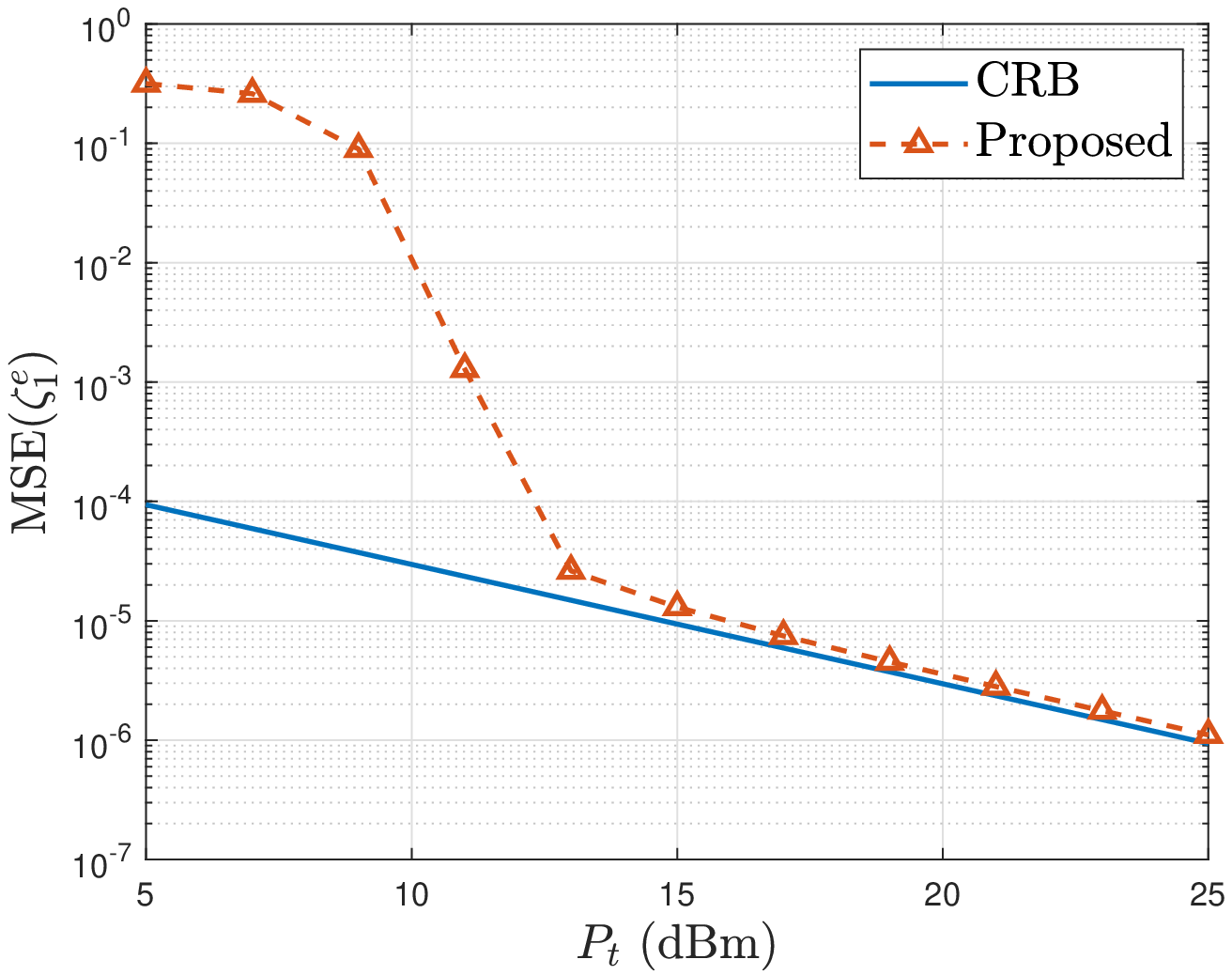}
  \end{minipage}
  \label{FIG6_5}
  }
    \subfigure[MSE and CRB of $\eta_1^e$.]{
  \begin{minipage}{.31\textwidth}
   \centering
   \includegraphics[scale=.38]{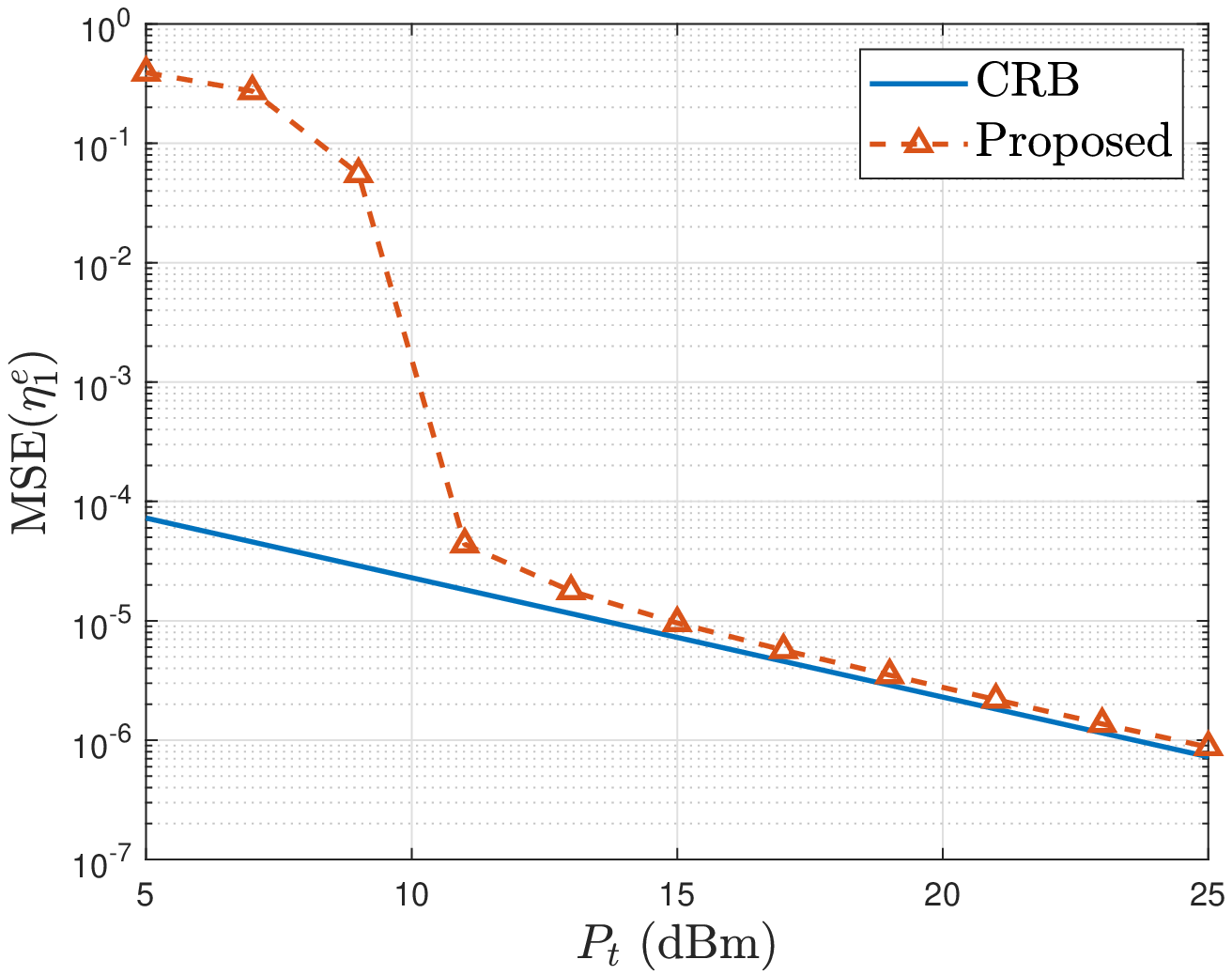}
  \end{minipage}
  \label{FIG6_6}
  }
  \hfill
      \subfigure[MSE and CRB of $\mathbf{p}_{\rm I}$.]{
  \begin{minipage}{.31\textwidth}
   \centering
   \includegraphics[scale=.38]{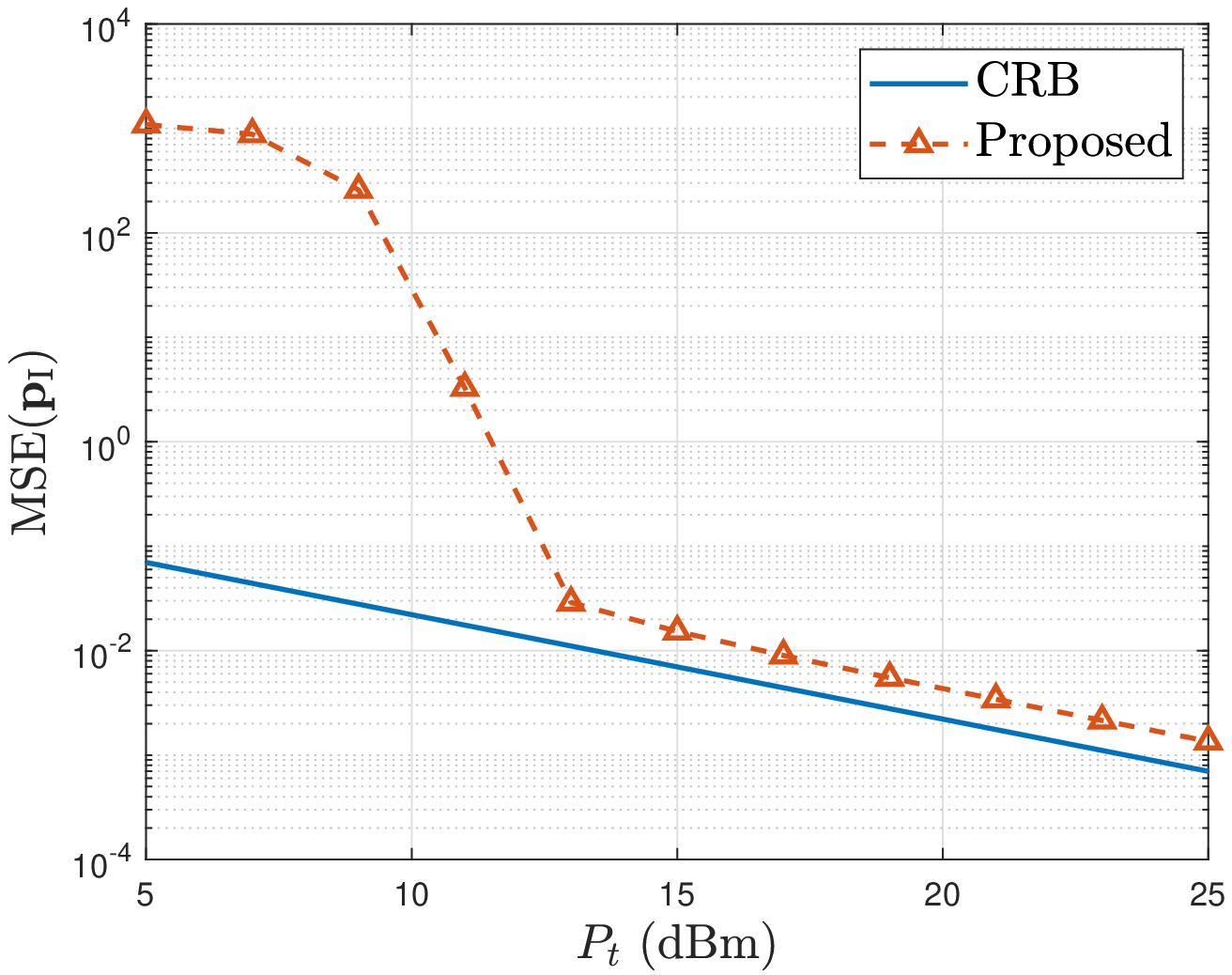}
  \end{minipage}
  \label{FIG6_7}
  }
    \subfigure[MSE and CRB of $\mathbf{Q}$.]{
  \begin{minipage}{.31\textwidth}
   \centering
   \includegraphics[scale=.38]{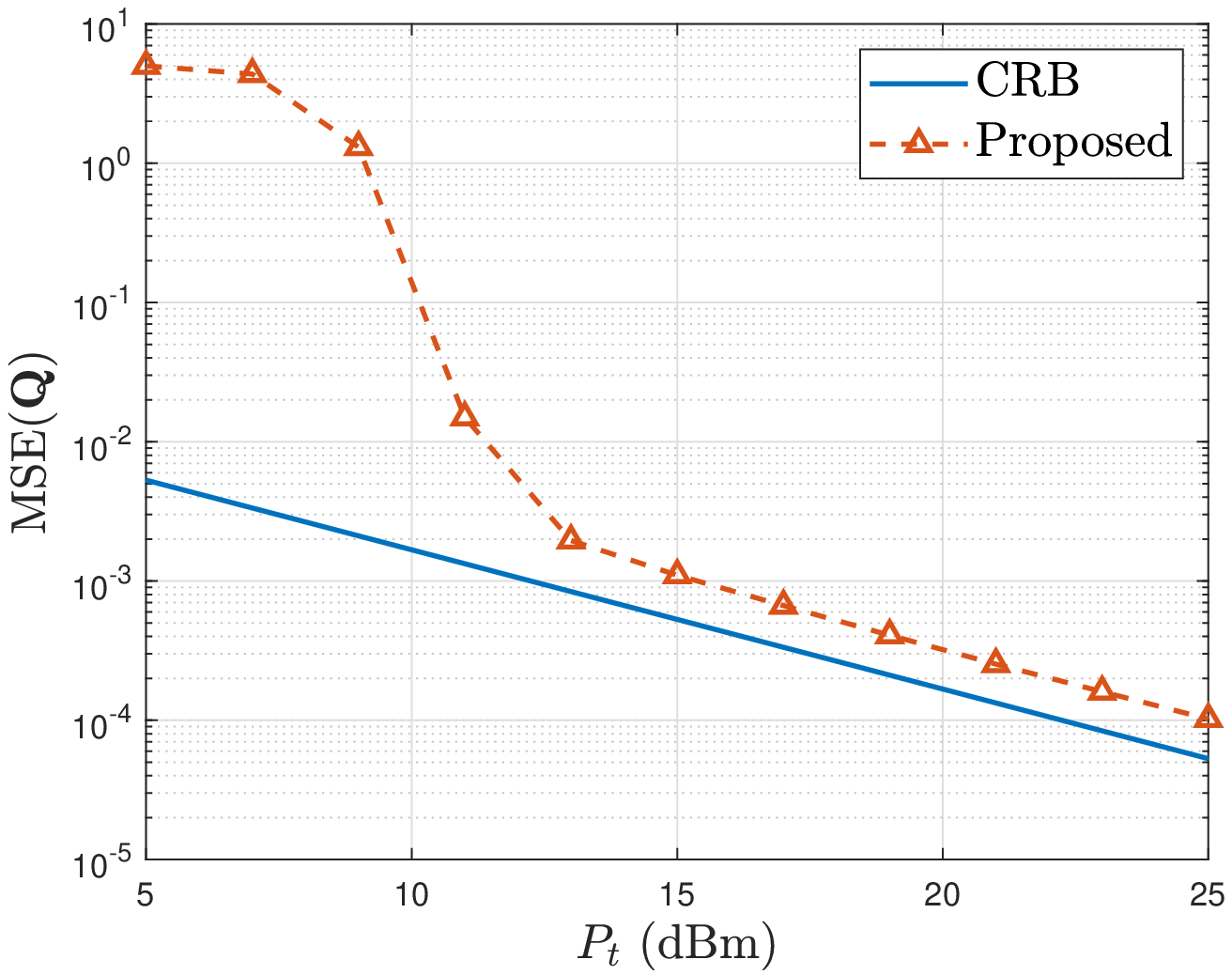}
  \end{minipage}
  \label{FIG6_8}
 }
 \caption{MSEs and CRBs of angle information acquisition and 6D information acquisition}
 \label{Fig-CRB}
\end{figure}

\subsection{CRBs and Sensitivity Analysis}
We first evaluate the estimation accuracy of the angle parameters associated with the TX, IRS, and RX $1$, i.e., $ \mathcal{A} =\{  \zeta_0^a, \zeta_1^a,\eta_1^a, \zeta_0^e, \zeta_1^e, \eta_1^e\}$ and the 6D information $\mathcal{Q} = \{\mathbf{p}_{\rm I} , \mathbf{Q} \}$ by the proposed algorithms, for which the mean square error (MSE) is utilized as a metric, which is given by ${\rm MSE}( x) = \mathbb E [ \| x - \hat{x} \|_2^2]$,
where $\hat{x}$ denote the estimate of $x$.

\begin{figure}[!t]
\centering
\begin{minipage}[t]{0.48\textwidth}
\centering
\includegraphics[scale=.5]{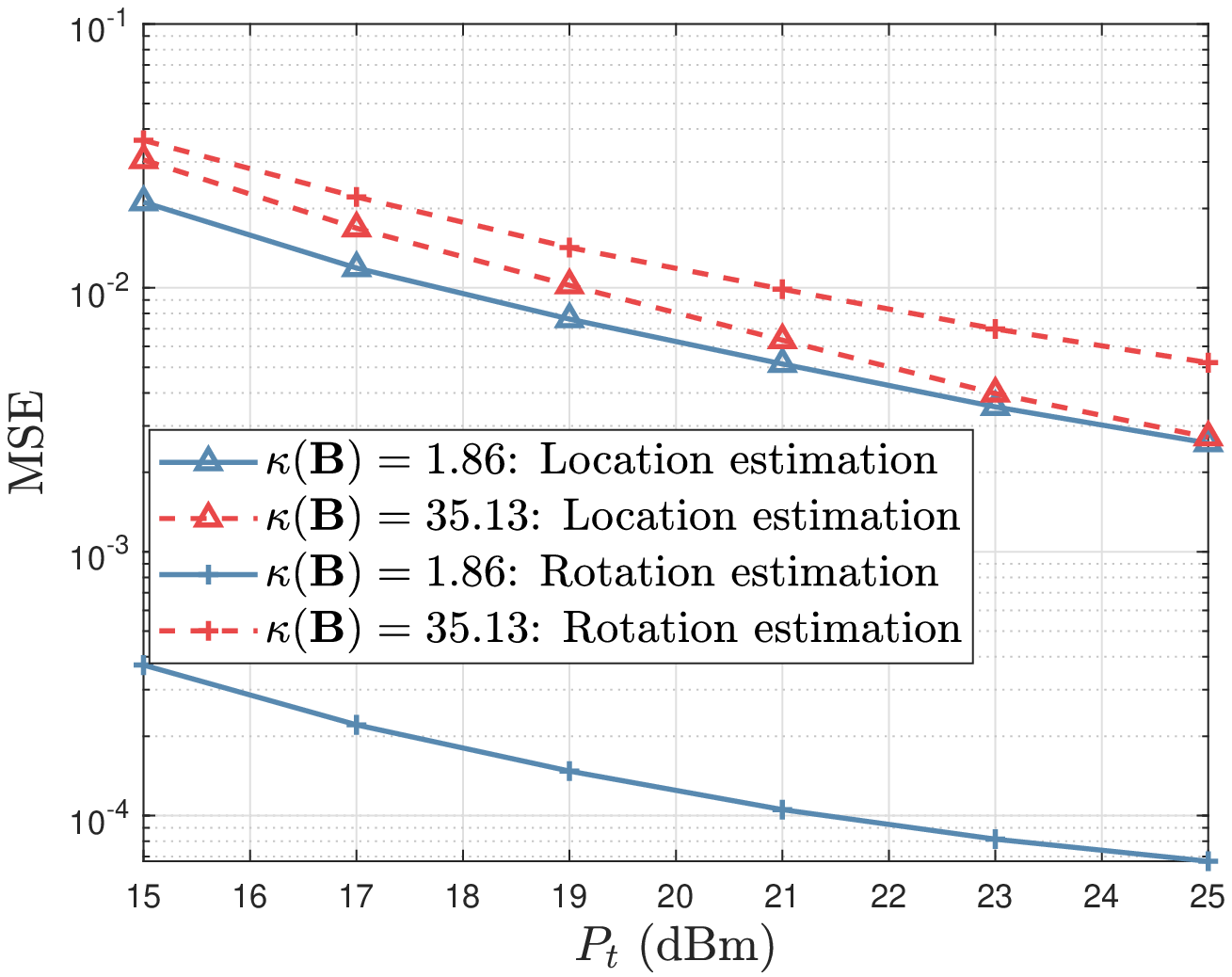}
\caption{MSEs of $\mathbf{p}_{\rm I}$ and $\mathbf{Q}$ versus $P_t$ under different values of $\kappa(\boldsymbol{B})$.}
\label{Fig_MSE_cond_fig}
\end{minipage}
\begin{minipage}[t]{0.48\textwidth}
\centering
\includegraphics[scale=.5]{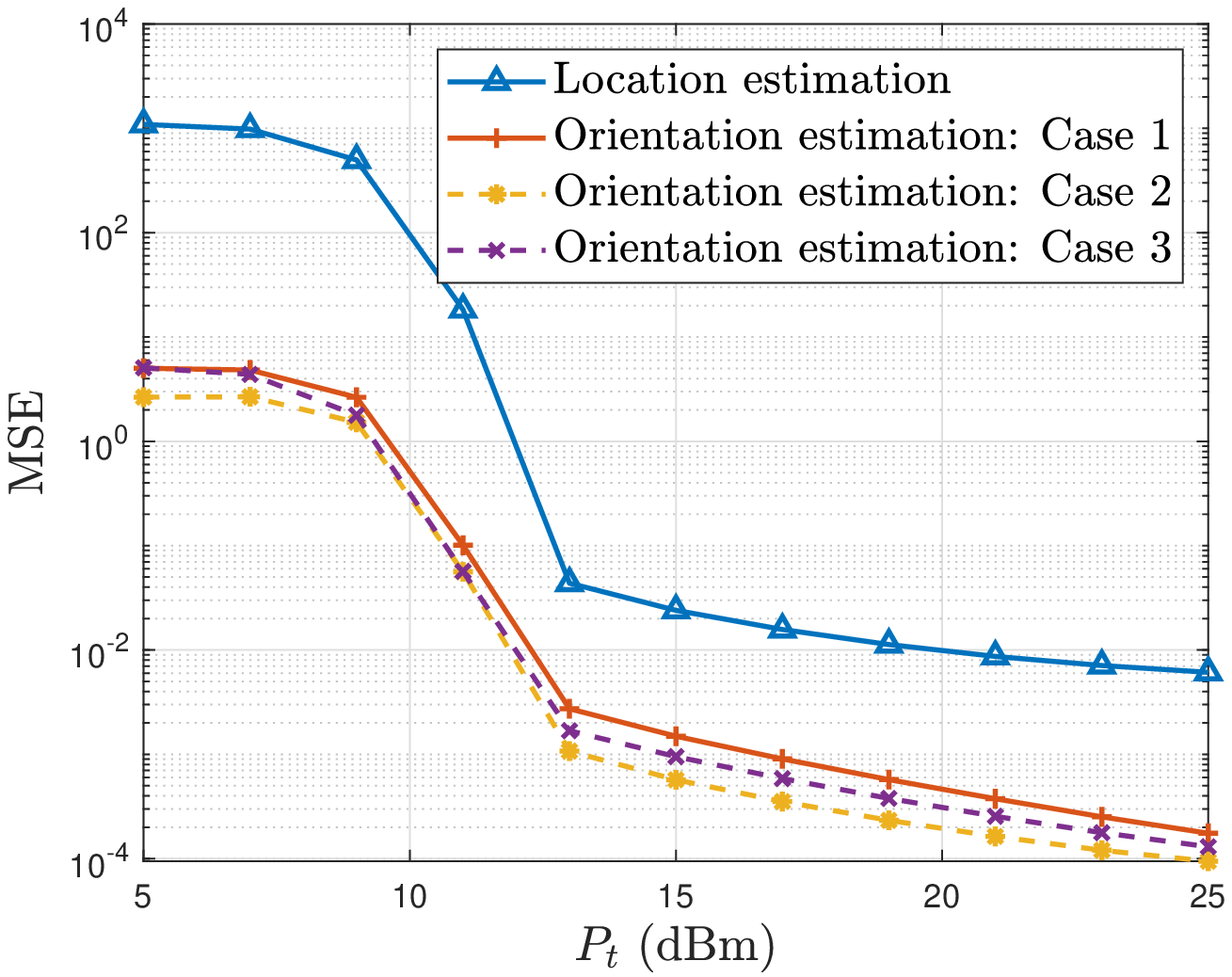}
\caption{MSEs of $\mathbf{p}_{\rm I}$ and $\mathbf{Q}$ versus $P_t$.}
\label{Fig_MSE_fig}
\end{minipage}
\end{figure}
In Figs. \ref{Fig-CRB}(a)-\ref{Fig-CRB}(f), we evaluate the MSE performance and CRBs on the angle parameters in $\mathcal{A}$ versus the transmit power $P_t$, respectively, where the Rician factors of all related channels are set to infinity to better evaluate the accuracy of the proposed estimators. It is observed that the CRBs on all parameters decrease exponentially while increasing the transmit power. In particular, the MSEs by the proposed tensor-based method are observed to be close to the CRBs in the high transmit power regime, which thus validates its efficacy. Furthermore, in Figs. \ref{Fig-CRB}(g)-\ref{Fig-CRB}(h), we plot the MSEs by the proposed 6D information acquisition scheme and the CRBs on the IRS's location $\mathbf{p}_{\rm I}$ and orientation $\mathbf{Q}$ (see \eqref{eq-PEB} and \eqref{eq-OEB}) versus the transmit power $P_t$. It is observed that the achieved MSEs decrease with $P_t$, similarly as the observations made in Figs. \ref{Fig-CRB}(a)-\ref{Fig-CRB}(f) for angle parameters. It is also observed that the performance gap between the achieved MSE and CRB on $\mathbf{Q}$ in Fig. \ref{Fig-CRB}(h) is larger than that on $\mathbf{p}_{\rm I}$ in Fig. \ref{Fig-CRB}(g). This is because the orientation estimation relies on both the estimated location $\mathbf{\hat p}_{\rm I}$ and angle parameters $\{\hat{\eta}_k^e, \hat{\eta}_k^a \}_{k=1}^K$, which may be subject to accumulated estimation error as compared to the location estimation.
 
Next, to verify the sensitivity results in Section \ref{sec-sense2}, we evaluate the effect of the condition number $\kappa(\boldsymbol{B}) $ on the accuracy of orientation estimation, where we set $K=3$ with $\mathbf{p}_{\rm RX,3} =[-46,6,10]^T$, and consider two different IRS locations, namely, $\mathbf{p}_{\rm I} =[5,4,10]^T$ and $\mathbf{p}_{\rm I} =[-5,-10,4]^T$. With these two IRS locations, we have $\kappa(\boldsymbol{B}) = 1.86$ and $35.13 $, respectively.
Fig. \ref{Fig_MSE_cond_fig} plots the MSEs of location estimation and orientation estimation versus the transmit power $P_t$ under the above two IRS locations. It is observed that although the performance of location estimation is similar under these two locations, the performance of orientation estimation shows significant differences. In particular, the accuracy of orientation estimation under $\kappa(\boldsymbol{B}) = 1.86 $ is observed to be much higher than that under $\kappa(\boldsymbol{B}) = 35.13 $. This implies that $\kappa(\boldsymbol{B})$ or the location of all nodes can dramatically affect the orientation estimation even if with a small MSE in location estimation, which is consistent with our theoretical analysis in Section \ref{sec-sense2}.

\subsection{Performance of 6D Information Acquisition}
 Since the orientation estimation relies on both the estimated location $\mathbf{\hat p}_{\rm I}$ and angle parameters $\{\hat{\eta}_k^e, \hat{\eta}_k^a \}_{k=1}^K$, to analyze their respective effects on the accuracy of orientation estimation, we compare the MSEs of orientation estimation in the following three cases, i.e., 
\begin{itemize}
\item Case 1: orientation estimation with estimated location and angle parameters ($\mathbf{\hat p}_{\rm I}$ and $\{\hat{\eta}_k^e, \hat{\eta}_k^a \}_{k=1}^K$), 
\item Case 2: orientation estimation with estimated/perfect location information/angle parameters ($\mathbf{\hat p}_{\rm I}$ and $\{{\eta}_k^e, {\eta}_k^a \}_{k=1}^K$),
\item  Case 3: orientation estimation with perfect/estimated location information/angle parameters ($\mathbf{ p}_{\rm I}$ and $\{\hat{\eta}_k^e, \hat{\eta}_k^a \}_{k=1}^K$).
\end{itemize}
 
In Fig. \ref{Fig_MSE_fig}, we plot the MSEs of location and orientation estimation under the above three cases versus the transmit power $P_t$ with $M_x =M_y =8$. It is interesting to observe that the MSE in Case 2 is lower than that in Case 3, which implies that the orientation estimation is more sensitive to the estimation error of the angle parameters $\{{\eta}_k^e, {\eta}_k^a \}_{k=1}^K$ than that of the location $\mathbf{p}_{\rm I}$, which validates the analysis at the end of Section \ref{sec-sense2}. An intuitive explanation is that even a moderate deviation from the actual IRS location $\mathbf{p}_{\rm I}$ (e.g., several meters) is much smaller as compared to the TX-IRS and IRS-RX $k$ distances, thus marginally affecting the orientation estimation; in contrast, a small deviation from the true angle parameters $\{{\eta}_k^e, {\eta}_k^a \}_{k=1}^K$ can significantly alter the estimate of the rotation matrix.
  
  \begin{figure}[!t]
\centering
\begin{minipage}[t]{0.48\textwidth}
\centering
\includegraphics[scale=.5]{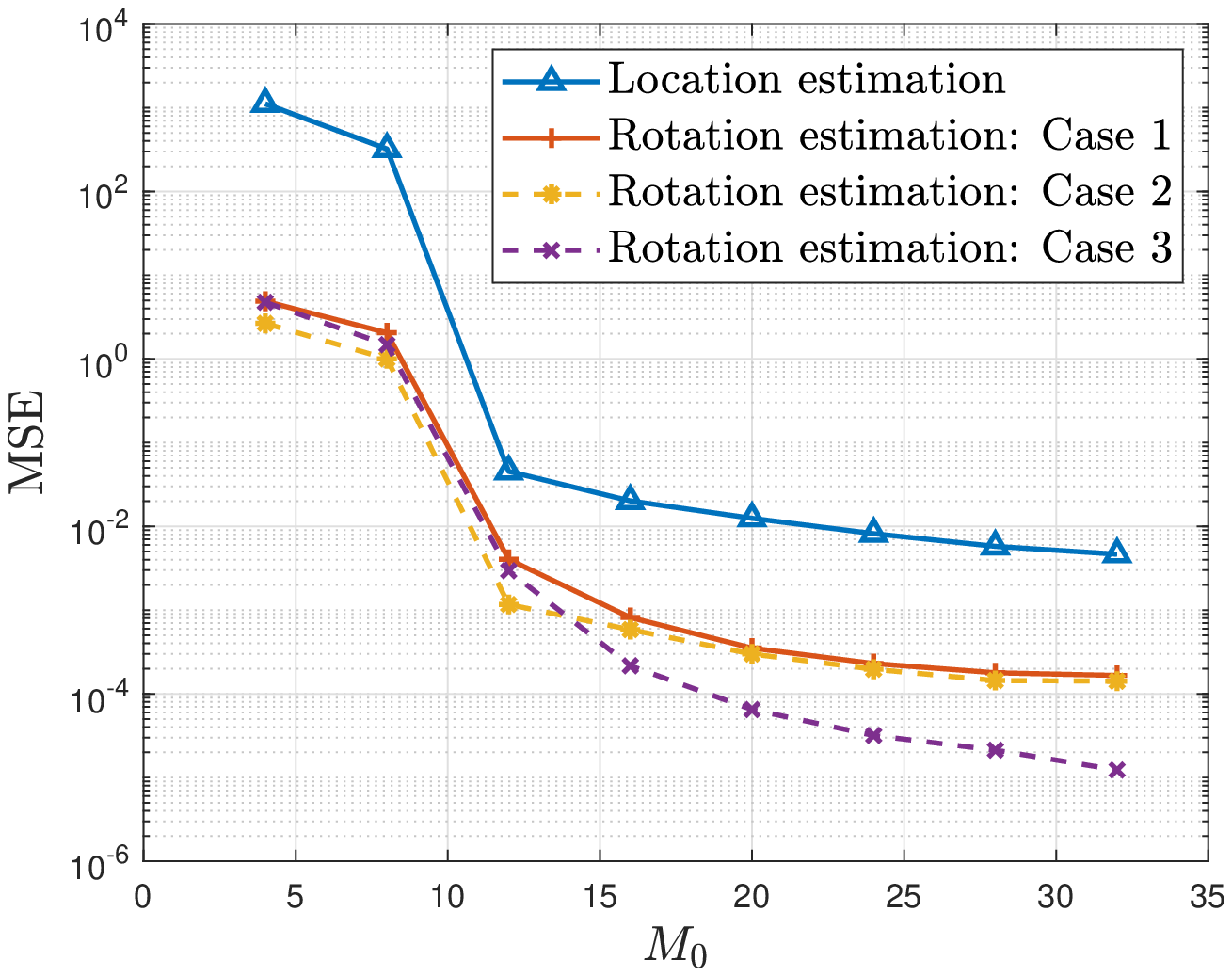}
\caption{MSEs of $\mathbf{p}_{\rm I}$ and $\mathbf{Q}$ versus $M_0$.}
\label{Fig_MSE_M_fig}
\end{minipage}
\begin{minipage}[t]{0.48\textwidth}
\centering
\includegraphics[scale=.5]{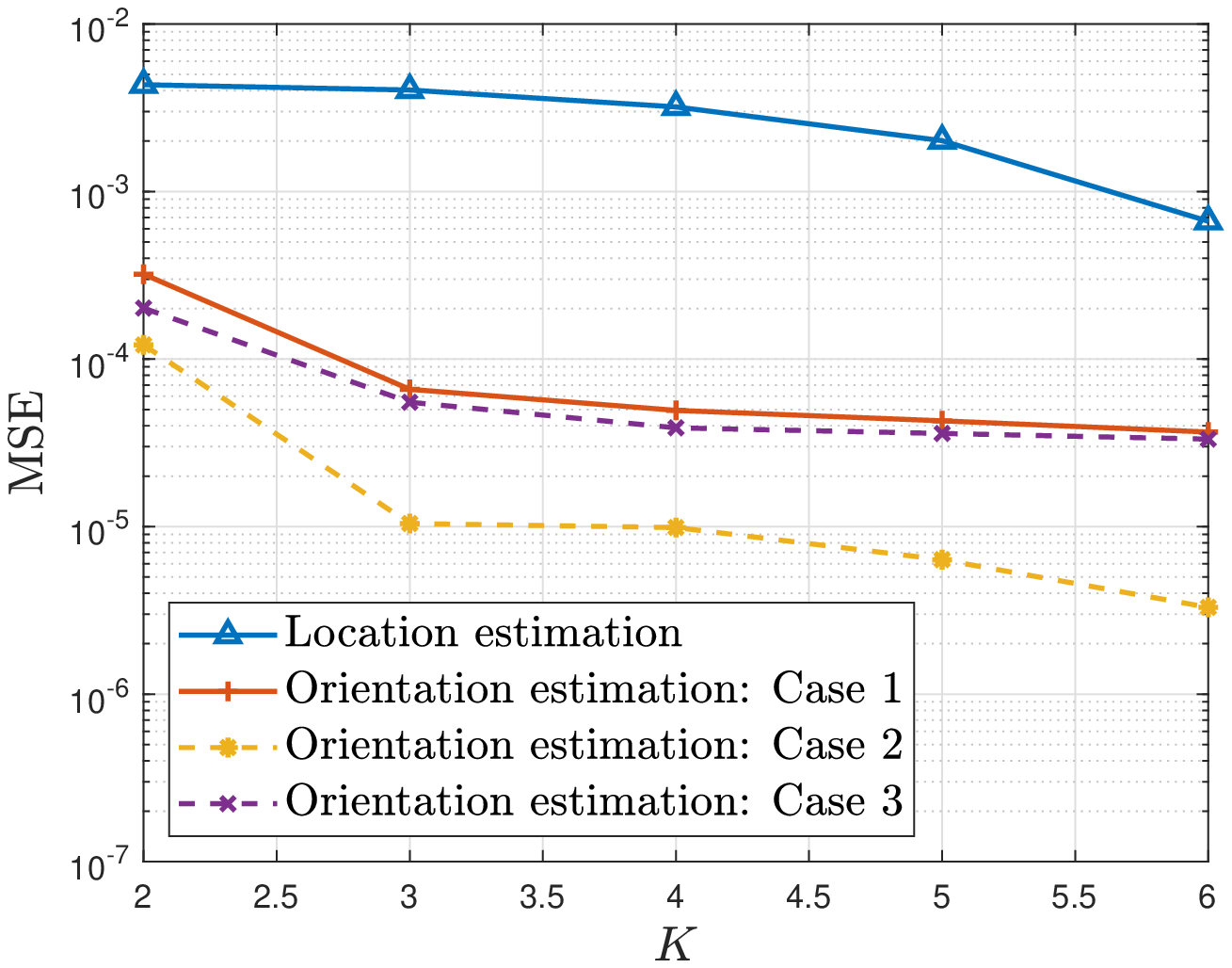}
\caption{MSEs of $\mathbf{p}_{\rm I}$ and $\mathbf{Q}$ versus $K$.}
\label{Fig_MSE_K_fig}
\end{minipage}
\end{figure}

Next, Fig. \ref{Fig_MSE_M_fig} shows the performance of the proposed methods versus the number of IRS reflecting elements in each dimension, with $M_x=M_y \triangleq M_0$. It is observed that all MSEs decrease monotonically with increasing $M_0$. This is because increasing $M_0$ helps boost the received signal power at each RX, thus improving the performance of angle estimation. Moreover, it can be seen that the MSE in Case 2 approaches that in Case 1 as $M_0$ increases. The reason is that increasing $M_0$ helps improve the estimation accuracy of the cascaded spatial frequencies $\{{\eta}_k^e, {\eta}_k^a \}_{k=1}^K$. Moreover, the MSE in Case 3 becomes lower than that in Case 2 when $M_0$ increases, which implies that the location estimation error may become a bottleneck to improve the orientation estimation accuracy even if the angle parameters $\{{\eta}_k^e, {\eta}_k^a \}_{k=1}^K$ are accurately estimated with increasing $M_0$.

Finally, in Fig. \ref{Fig_MSE_K_fig}, we show the MSEs of the location and orientation estimation versus the number of RXs, $K$, with $P_t = 20$ dBm. The locations of RXs 3-6 are set to $\mathbf{p}_{\rm RX, 3} = [-46,6,10]^T$, $\mathbf{p}_{\rm RX, 4} = [-2,38,3]^T$, $\mathbf{p}_{\rm RX, 5} = [20,-17,19]^T$, and $\mathbf{p}_{\rm RX, 6} = [5,10,11]^T$, respectively. It is observed that the MSE of location estimation decreases with $K$, thanks to the more observations collected. Moreover, although the MSE of orientation estimation decreases with $K$ over the whole range in Cases 2, it is observed to be approximately constant when $K>3$ in Cases 1 and 3. This implies that in contrast to the location estimation, the accuracy of angle parameter estimation may not improve as $K$ increases. This is because each RX $k$ \emph{independently} estimates its angle parameters ${\eta}_k^e$ and ${\eta}_k^a$, and hence the resulting angle estimation accuracy is regardless of the total number of RXs.

\section{Conclusion}
\label{sec-conclu}
This paper proposes a new device-free sensing system for joint location and orientation estimation enabled by a target-mounted IRS. To this end, we first propose a tensor-based method to acquire essential angle information between the IRS and the sensing TX, as well as a set of sensing RXs, based on which the location and orientation of the IRS/target are efficiently estimated by solving two least-square problems, respectively. Theoretical analysis is provided to derive the CRBs on the involved parameters and unveil the non-trivial impacts of the locations of the sensing TX and RXs, as well as the estimation accuracy of the IRS's cascaded spatial frequencies on the orientation estimation accuracy. In particular, the orientation estimation accuracy is shown to be very robust against the estimation error in the IRS's location. Simulation results validate the above analytic results and demonstrate the efficacy of the proposed sensing scheme even when the number of sensing RXs is small. There are several promising directions worthy of further investigation for the proposed target-mounted IRS-assisted sensing in future work, such as its applications for multi-target sensing, target recognition and tracking, integrated sensing and communication (ISAC), etc.

\useRomanappendicesfalse
\appendices

\section{Calculation of FIM $\boldsymbol{I}(\boldsymbol{\Gamma})$}
\label{AppendixA} 

Based on \eqref{eq-Yk}, the log-likelihood function of the parameter vector $\boldsymbol{\Gamma}$ in \eqref{eq-Gamma} can be expressed as 
\begin{align}
\mathcal{L}(\boldsymbol{\Gamma}) =&- \tilde{D}  - \frac{1}{\sigma^2}  \sum_{k=1}^K
\| \boldsymbol{Y}_{k,(1)} - \boldsymbol{\tilde a}_{r,k} (\boldsymbol{\tilde a}_{I,k}^{ } \odot \boldsymbol{\tilde a}_t^{} )^T\|_F^2, \nonumber \\
=& -\tilde{D}-\frac{1}{\sigma^2} \sum_{k=1}^K \| \boldsymbol{Y}_{k,(2)} - \boldsymbol{\tilde a}_t (\boldsymbol{\tilde a}_{I,k}^{ }  \odot \boldsymbol{\tilde a}_{r,k} ^{ } )^T\|_F^2, \nonumber \\
=& -\tilde{D}-\frac{1}{\sigma^2} \sum_{k=1}^K \| \boldsymbol{Y}_{k,(3)} - \boldsymbol{\tilde a}_{I,k} (\boldsymbol{\tilde a}_t^{ } \odot \boldsymbol{\tilde a}_{r,k} ^{ })^T\|_F^2 ,
\end{align}
where $\tilde{D} \triangleq D_{\rm Re} D_{\rm BS} D_{\rm IRS} K \ln \pi \sigma^2 $.
Then, the FIM for $\boldsymbol{\Gamma}$ is given by $
\boldsymbol{I}(\boldsymbol{\Gamma}) = \mathbb E \left \{  \left( \frac{\partial \mathcal{L}(\boldsymbol{\Gamma})}{\partial \boldsymbol{\Gamma} } \right)^T  \frac{\partial \mathcal{L}(\boldsymbol{\Gamma})}{\partial \boldsymbol{\Gamma} }  \right\}$\cite{Kay93,ZhouFang17}. To calculate $\boldsymbol{I}(\boldsymbol{\Gamma})$, we first compute the partial derivative of $\mathcal{L}(\boldsymbol{\Gamma})$ w.r.t. $\boldsymbol{\Gamma}$ and then calculate its expectation, as presented in the subsequent two subsections, respectively.

\subsection{Partial Derivatives of $\mathcal{L}(\boldsymbol {\Gamma})$ w.r.t. $\boldsymbol{\Gamma}$ }
First, the partial derivative of $\mathcal{L}(\boldsymbol{\Gamma})$ w.r.t. $\zeta_0^a$ can be calculated as
\begin{align}
\frac{\partial \mathcal{L}(\boldsymbol{\Gamma})}{\partial \zeta_0^a } = {\rm tr} \left \{  \left( \frac{\partial \mathcal{L}(\boldsymbol{\Gamma})}{\partial \boldsymbol{\tilde a}_t } \right)^T  \frac{\partial  \boldsymbol{\tilde a}_t}{\partial \zeta_0^a }  + \left( \frac{\partial \mathcal{L}(\boldsymbol{\Gamma})}{\partial \boldsymbol{\tilde a}_t^{\ast} } \right)^T \frac{\partial  \boldsymbol{\tilde a}_t^{\ast}}{\partial \zeta_0^a }  \right \} ,
\label{eq-partial-g}
\end{align}
where
\begin{align}
&\frac{\partial \mathcal{L}(\boldsymbol{\Gamma})}{\partial \boldsymbol{\tilde a}_t} =  \left( \frac{\partial \mathcal{L}(\boldsymbol{\Gamma})}{\partial \boldsymbol{\tilde a}_t^{\ast} } \right)^{\ast}  = \frac{1}{\sigma^2} \sum_{k=1}^K\left( \boldsymbol{Y}_{k,(2)}^T -  (\boldsymbol{\tilde a}_{I,k}^{ }  \odot \boldsymbol{\tilde a}_{r,k}^{ } ) \boldsymbol{\tilde a}_t ^T\right)^H 
\left( \boldsymbol{\tilde a}_{I,k}^{ }  \odot \boldsymbol{\tilde a}_{r,k}^{ } \right)   \nonumber \\
&\frac{\partial \boldsymbol{\tilde a}_t }{ \partial \zeta_0^a} =\left(\frac{\partial \boldsymbol{\tilde a}_t ^{\ast}}{ \partial \zeta_0^a} \right)^{\ast} = -j\boldsymbol{F}^{T}\boldsymbol{D}_a (N_{t_y},N_{t_z}) \boldsymbol{a}_t^{\ast}(\zeta_0^e, \zeta_0^a) , \nonumber \\
&\boldsymbol{D}_a(N_1,N_2) =\pi {\rm diag} \big(    [0,1,\ldots, N_1-1]^T \otimes  \boldsymbol{1}_{N_2}\big).
\end{align}
By performing some manipulations on \eqref{eq-partial-g}, it can be simplified as $\frac{\partial \mathcal{L}(\boldsymbol{\Gamma})}{\partial \zeta_0^a } =  \frac{2}{\sigma^2}  \sum_{k=1}^K \Re \{     ( \boldsymbol{\tilde a}_{I,k}^{ }  \odot \boldsymbol{\tilde a}_{r,k}^{ } )^T ( \boldsymbol{Y}_{k,(2) }^T -  (\boldsymbol{\tilde a}_{I,k}^{ }  \odot \boldsymbol{\tilde a}_{r,k}^{ } ) \boldsymbol{\tilde a}_t ^T)^{\ast} \frac{\partial \boldsymbol{\tilde a}_t }{ \partial \zeta_0^a} \}$.
Similarly, we can obtain the partial derivatives w.r.t. the other angle parameters as follows,
\begin{align}
&\frac{\partial \mathcal{L}(\boldsymbol{\Gamma})}{\partial \zeta_0^e } =\frac{2}{\sigma^2}  \sum_{k=1}^K \Re \left \{     \left( \boldsymbol{\tilde a}_{I,k}^{ }  \odot \boldsymbol{\tilde a}_{r,k}^{ } \right)^T \left( \boldsymbol{Y}_{k,(2) }^T -  (\boldsymbol{\tilde a}_{I,k}^{ }  \odot \boldsymbol{\tilde a}_{r,k}^{ } ) \boldsymbol{\tilde a}_t ^T\right)^{\ast} \frac{\partial \boldsymbol{\tilde a}_t }{ \partial \zeta_0^e} \right \}, \\
&\frac{\partial \mathcal{L}(\boldsymbol{\Gamma})}{\partial \zeta_k^a } = \frac{2}{\sigma^2} \Re \left \{ \left( \boldsymbol{\tilde a}_{I,k}^{ }  \odot \boldsymbol{\tilde a}_t^{ }  \right)^T \left( \boldsymbol{Y}_{k,(1)}^T -  (\boldsymbol{\tilde a}_{I,k}^{ }  \odot \boldsymbol{\tilde a}_t^{ } ) \boldsymbol{\tilde a}_{r,k}^T\right)^{\ast} \frac{\partial \boldsymbol{\tilde a}_{r,k} }{ \partial \zeta^a} \right \},\\
&\frac{\partial \mathcal{L}(\boldsymbol{\Gamma})}{\partial \zeta_k^e } = \frac{2}{\sigma^2} \Re \left \{ \left( \boldsymbol{\tilde a}_{I,k}^{ }  \odot \boldsymbol{\tilde a}_t^{ }  \right)^T \left( \boldsymbol{Y}_{k,(1)}^T -  (\boldsymbol{\tilde a}_{I,k}^{ }  \odot \boldsymbol{\tilde a}_t^{ } ) \boldsymbol{\tilde a}_{r,k}^T\right)^{\ast} \frac{\partial \boldsymbol{\tilde a}_{r,k} }{ \partial \zeta^a} \right \},\\
&\frac{\partial \mathcal{L}(\boldsymbol{\Gamma})}{\partial \eta_k^a } =\frac{2}{\sigma^2} \Re \left \{ \left( \boldsymbol{\tilde a}_t^{ }  \odot \boldsymbol{\tilde a}_{r,k}^{ }  \right)^T \left( \boldsymbol{Y}_{k,(3)}^T -  (\boldsymbol{\tilde a}_t^{ }  \odot \boldsymbol{\tilde a}_{r,k}^{ } ) \boldsymbol{\tilde a}_{I,k}^T\right)^{\ast} \frac{\partial \boldsymbol{\tilde a}_{I,k} }{ \partial \eta_k^a} \right \},\\
&\frac{\partial \mathcal{L}(\boldsymbol{\Gamma})}{\partial \eta_k^e } =\frac{2}{\sigma^2} \Re \left \{ \left( \boldsymbol{\tilde a}_t^{ }  \odot \boldsymbol{\tilde a}_{r,k}^{ }  \right)^T \left( \boldsymbol{Y}_{k,(3)}^T -  (\boldsymbol{\tilde a}_t^{ }  \odot \boldsymbol{\tilde a}_{r,k}^{ } ) \boldsymbol{\tilde a}_{I,k}^T\right)^{\ast} \frac{\partial \boldsymbol{\tilde a}_{I,k} }{ \partial \eta_k^e} \right \},
\end{align}
where
\begin{align}
\frac{\partial \boldsymbol{\tilde a}_t }{ \partial \zeta_0^e} = &  \left( -j\boldsymbol{F}^{T}\boldsymbol{D}_e(N_{t_x},N_{t_y}) \boldsymbol{a}_t^{\ast}(\zeta_0^e, \zeta_0^a) \right), 
\frac{\partial \boldsymbol{\tilde a}_{r,k} }{ \partial \zeta_k^a}  = j \boldsymbol{W}_k^H \boldsymbol{D}_a(N_{r_y},N_{r_z}) \boldsymbol{a}_{r} ({\zeta}_k^e,\zeta_k^a) ,\nonumber \\
\frac{\partial \boldsymbol{\tilde a}_{r,k} }{ \partial \zeta_k^e}  =& j \boldsymbol{W}_k^H \boldsymbol{D}_e(N_{r_y},N_{r_z}) \boldsymbol{a}_{r} ({\zeta}_k^e,\zeta_k^a),
\frac{\partial \boldsymbol{\tilde a}_{r,k} }{ \partial \eta_k^a}  = j \boldsymbol{V}^H \boldsymbol{D}_a (M_x,M_y)\boldsymbol{a}_{I} ({\eta}_k^e,\eta_k^a) ,\nonumber\\
\frac{\partial \boldsymbol{\tilde a}_{r,k} }{ \partial \eta_k^e}  =& j \boldsymbol{V}^H \boldsymbol{D}_e (M_x,M_y)\boldsymbol{a}_{I} ({\eta}_k^e,\eta_k^a) , \boldsymbol{D}_e(N_1,N_2) = \pi {\rm diag}( \boldsymbol{1}_{N_1} \otimes [0,1, \ldots, N_2-1]^T ).
\end{align}
\subsection{Calculation of FIM $\boldsymbol{I}(\boldsymbol{\Gamma})$}
To calculate the FIM $\boldsymbol{I}(\boldsymbol{\Gamma})$, we start from its diagonal and subdiagonal entries.
Due to the tedious calculations for all these entries and space limit, we only show the details for some of them. For example, the $(1,1)$-th or the first diagonal element of $\boldsymbol{I}(\boldsymbol{\Gamma})$ is given by $\mathbb E \left\{ \left( \frac{\partial \mathcal{L}(\boldsymbol{\Gamma})}{\partial \zeta_0^a }  \right)^{\ast} \frac{\partial \mathcal{L}(\boldsymbol{\Gamma})}{\partial \zeta_0^a }        \right\} = 4  \mathbb E \left[      \Re \{\bar c_{0,a} \}^2 \right]$,
where
\begin{align}
 \bar c_{0,a}\triangleq &\sum_{k=1}^K\frac{1}{\sigma^2}\left( \boldsymbol{\tilde a}_{I,k}^{ }  \odot \boldsymbol{\tilde a}_{r,k}^{ } \right)^T \left( \boldsymbol{Y}_{k,(2) }^T -  (\boldsymbol{\tilde a}_{I,k}^{ }  \odot \boldsymbol{\tilde a}_{r,k}^{ } ) \boldsymbol{\tilde a}_t^T\right)^{\ast} \frac{\partial \boldsymbol{\tilde a}_t }{ \partial \zeta_0^a}= \frac{1}{\sigma^2}\sum_{k=1}^K\left( \boldsymbol{\tilde a}_{I,k}^{ }  \odot \boldsymbol{\tilde a}_{r,k}^{ } \right)^T \boldsymbol{N}_{k,(2)}^H \frac{\partial \boldsymbol{\tilde a}_t }{ \partial \zeta_0^a}\nonumber \\
=& \frac{1}{\sigma^2} \sum_{k=1}^K \left( \left( \frac{\partial \boldsymbol{\tilde a}_t }{ \partial \zeta_0^a}\right)^T \otimes \left( \boldsymbol{\tilde a}_{I,k}^{ }  \odot \boldsymbol{\tilde a}_{r,k}^{ } \right)^T  \right) {\rm vec}\left( \boldsymbol{N}_{k,(2)}^H \right) = \frac{1}{\sigma^2} \sum_{k=1}^K \boldsymbol{\bar u}_{k,a}^T \boldsymbol{n}_{k,(2)},
\label{eq-ca}
\end{align} with $\boldsymbol{\bar u}_{k,a}^T  \triangleq   \left( \frac{\partial \boldsymbol{\tilde a}_t }{ \partial \zeta_0^a}\right)^T \otimes ( \boldsymbol{\tilde a}_{I,k}^{ }  \odot \boldsymbol{\tilde a}_{r,k}^{ } )^T$ and $
\boldsymbol{n}_{k,(2)} \triangleq   {\rm vec}\left( \boldsymbol{N}_{k,(2)}^H \right)$. 
It follows from \eqref{eq-ca} that $c_{0,a}$ is the linear transformation of $ \boldsymbol{n}_{k,(2)}$ and follows the CSCG distribution. As the effective received noises at different RXs are mutually independent, their variance and second-order moments are given by $\mathbb{E} ( \bar c_{0,a}  \bar c_{0,a}^{\ast}) = \frac{1}{\sigma^2}  \sum_{k=1}^K \boldsymbol{\bar u}_{k,a}^T \boldsymbol{\bar u}_{k,a}^{\ast}$ and $  \mathbb{E} ( \bar c_{0,a}  \bar c_{0,a}) = 0
\label{eq-cE}$, respectively, which result in
\begin{align}
\mathbb E \left\{ \left( \frac{\partial \mathcal{L}(\boldsymbol{\Gamma})}{\partial \zeta_0^a }  \right)^{\ast} \frac{\partial \mathcal{L}(\boldsymbol{\Gamma})}{\partial \zeta_0^a }        \right\} =  \frac{2}{\sigma^2}  \sum_{k=1}^K \boldsymbol{\bar u}_{k,a}^T \boldsymbol{\bar u}_{k,a}^{\ast} .
\label{eq-cE}
\end{align}
For the $(1,2)$-th or the first subdiagonal element of $\boldsymbol{I}(\boldsymbol{\Gamma})$, it can be calculated as $\mathbb E \big\{ ( \frac{\partial \mathcal{L}(\boldsymbol{\Gamma})}{\partial \zeta_0^a }  )^{\ast} \frac{\partial \mathcal{L}(\boldsymbol{\Gamma})}{\partial \zeta_0^e }        \big\} =  \frac{2}{\sigma^2}  \sum_{k=1}^K \Re \big( \boldsymbol{\bar u}_{k,a}^T \boldsymbol{\bar u}_{k,e}^{\ast} \big)$,
where $\boldsymbol{\bar u}_{k,e}^T  \triangleq  \left( \frac{\partial \boldsymbol{\tilde a}_t }{ \partial \zeta_0^e}\right)^T \otimes \left( \boldsymbol{\tilde a}_{I,k}^{ }  \odot \boldsymbol{\tilde a}_{r,k}^{ } \right)^T$. Following the same procedure, we can obtain all $(i,i)$-th (diagonal) and $(i,i+1)$-th (subdiagonal) elements, for which the details are omitted.

Next, we derive the expressions of the off-diagonal entries of $\boldsymbol{I}(\boldsymbol{\Gamma})$. Similarly, we only show the expressions for partial off-diagonal entries. For example, the $(1,k+2)$-th element of $\boldsymbol{I}(\boldsymbol{\Gamma})$ is given by
\begin{align}
\mathbb E \left\{ \left( \frac{\partial \mathcal{L}(\boldsymbol{\Gamma})}{\partial \zeta_0^a }  \right)^{\ast} \frac{\partial \mathcal{L}(\boldsymbol{\Gamma})}{\partial \zeta_k^a }        \right\}  =  \mathbb E ( (\bar c_{0,a} + \bar c_{0,a}^{\ast}) (c_{k,a} + c_{k,a}^{\ast})),
\label{eq-21}
\end{align}where $\bar c_{0,a}$ is defined in \eqref{eq-ca} and $
c_{k,a} \triangleq \frac{1}{\sigma^2}  \boldsymbol{ u}_{k,a}^T \boldsymbol{n}_{k,(1)}$ with $ \boldsymbol{n}_{k,(1)} ={\rm vec}\left( \boldsymbol{N}_{k,(1)}^H \right)$. It follows from \eqref{eq-cE} that their second-order moments are given by $\mathbb E( \bar c_{0,a} c_{k,a}) = 0$, whilst their correlation can be calculated as
\begin{align}
\mathbb E(  \bar {c}_{0,a} c_{k,a}^{\ast}) = &\frac{1}{\sigma^4}   \mathbb E \left \{  \sum_{i=1}^K \boldsymbol{\bar u}_{i,a}^T\boldsymbol{n}_{i,(2)} \boldsymbol{n}_{k,(1)}^H \boldsymbol{ u}_{k,a}^{\ast}  \right \} 
\stackrel{(a)}=  \frac{1}{\sigma^4}   \boldsymbol{\bar u}_{k,a}^T \mathbb E(\boldsymbol{n}_{k,(2)} \boldsymbol{n}_{k,(1)}^H ) \boldsymbol{ u}_{k,a}^{\ast}= \frac{1}{\sigma^4}   \boldsymbol{\bar u}_{k,a}^T \boldsymbol{C}_{2,1}\boldsymbol{ u}_{k,a}^{\ast}, \nonumber
\end{align}
where in $(a)$ we use the property that the received noise at different RXs is independent of each other, and $\boldsymbol{C}_{2,1} \triangleq  E(\boldsymbol{n}_{k,(2)} \boldsymbol{n}_{k,(1)}^H ).$
Therefore, \eqref{eq-21} becomes $\mathbb E \{ ( \frac{\partial \mathcal{L}(\boldsymbol{\Gamma})}{\partial \zeta_0^a }  )^{\ast} \frac{\partial \mathcal{L}(\boldsymbol{\Gamma})}{\partial \zeta_k^a }  \}  =  \frac{2}{\sigma^4}   \Re \{ \boldsymbol{\bar u}_{k,a}^T \boldsymbol{C}_{2,1}\boldsymbol{ u}_{k,a}^{\ast} \}.$
Next, we compute $\boldsymbol{C}_{2,1} $ in the above fomula. Since the entries in $\boldsymbol{\mathcal{N}}_k $ are all i.i.d. Gaussian random variables, we have
\begin{align}
\mathbb E( n_{k, {i_1,j_1,q_1} }  n_{k,{i_2,j_2,q_2}  }^{\ast}) =   \begin{cases} \sigma^2, &  {\text{if } } i_1=i_2, j_1=j_2,q_1=q_2 ,\\
0,& {\text {otherwise}.}
\end{cases}
\end{align}
Hence, there should exist $D_{\rm RX}D_{\rm BS} D_{\rm IRS}$ nonzero entries in the correlation matrices $\boldsymbol{C}_{2,1}$, and its $(m,n)$-th element, based on the relationship between $\boldsymbol{{\mathcal N}}_k$ and $\boldsymbol{N}_{k,(1)}$/$\boldsymbol{N}_{k,(2)}$, is expressed as
\begin{align}
\boldsymbol{C}_{2,1} [m,n]=  \begin{cases} \sigma^2,    & {\text{if }} m =M_2(i,j,q)   , n =M_1(i,j,q) , \\
0,& {\text{otherwise,}}
\end{cases}
\end{align}
where 
$
M_1(i,j,q) =  j+ (q-1)D_{\rm TX} + (i-1)D_{\rm TX} D_{\rm IRS} $ and $
M_2(i,j,q)  = i + (q-1)D_{\rm RX} + (j-1)D_{\rm RX}D_{\rm IRS} . 
$ By performing similar operations as above, we can calculate the other off-diagonal elements of $\boldsymbol{I}(\boldsymbol{\Gamma })$, for which the details are omitted for brevity.


\section{Calculation of Jacobian Matrix $\nabla_{\boldsymbol{s}} \boldsymbol{\Gamma}$}
\label{AppendixB} 
The $(i,j)$-th element of $\nabla_{\boldsymbol{s}} \boldsymbol{\Gamma}$ is given by $\frac{\partial \Gamma[i]}{ \partial  s[j]}$, where $\Gamma[i]$ is the $i$-th entry of the vector $\boldsymbol{\Gamma}$ defined in \eqref{eq-Gamma} and $s[j]$ is the $j$-th entry of the vector $\mathbf{s} =[ \mathbf{p}_{\rm I}^T, \boldsymbol{\psi}^T]^T$. Note that the angle parameters $\boldsymbol{\zeta}^a$ and $\boldsymbol{\zeta}^e$ are regardless of the axis angles $\boldsymbol{\psi}$, and their partial derivatives w.r.t. the location $\mathbf{p}_{\rm I}$ have been provided in \eqref{eq-deri}. Next, we focus on the partial derivatives of angle parameters $\boldsymbol{\eta}^a$ and $\boldsymbol{\eta}^e$ w.r.t. $\mathbf{p}_{\rm I}$ and $\boldsymbol{\psi}$, respectively.

For clarity, we rewrite $\eta_k^a$ and $\eta_k^e$  in \eqref{eq-etaqa} and \eqref{eq-etaqe} as a function of $\mathbf{p}_{\rm I}$ and $\mathbf{Q}$, i.e., $\eta^a_k  ( \mathbf{p}_{\rm I}, \mathbf{Q} ) =   \boldsymbol{e}_1^T \mathbf{Q}^T \boldsymbol{b}_k(\mathbf{p}_{\rm I}) $, and $\eta^e_k ( \mathbf{p}_{\rm I}, \mathbf{Q} )  =   \boldsymbol{e}_2^T \mathbf{Q}^T \boldsymbol{b}_k (\mathbf{p}_{\rm I}) , \forall k  \label{eq-etaqe-2} $, where $\mathbf{b}_k(\mathbf{p}_{\rm I})$ is defined in \eqref{eq-bk}. Based on the above, the partial derivative of $\eta_k^a$ and $\eta_k^e$ w.r.t. $\mathbf{p}_{\rm I}$ are given by $
\frac{\partial \eta_k^a(\mathbf{p}, \mathbf{Q})}{\partial \mathbf{p}}  =  
\boldsymbol{J}_{b_k} (\mathbf{p}_{\rm I}) \mathbf{Q} \boldsymbol{e}_1 $, and $
\frac{\partial \eta_k^e(\mathbf{p}, \mathbf{Q})}{\partial \mathbf{p}}  =  
\boldsymbol{J}_{b_k} (\mathbf{p}_{\rm I}) \mathbf{Q} \boldsymbol{e}_2 $, respectively,
where $\boldsymbol{J}_{b_k} (\mathbf{p})$ is the Jacobian matrix of $\mathbf{b}_k(\mathbf{p}_{})$ w.r.t. $\mathbf{p}_{}$, i.e., $\boldsymbol{J}_{b_k} (\mathbf{p})= [ \frac{\partial \mathbf{b}_k[1]}{\partial \mathbf{p}}, \frac{\partial \mathbf{b}_k[2]}{\partial \mathbf{p}}, \frac{\partial \mathbf{b}_k[3]}{\partial \mathbf{p}} ]  \in \mathbb R^{3 \times 3}$, with $\boldsymbol{b}_k[i]$ being the $i$-th entry of $\boldsymbol{b}_k(\mathbf{p})$, i.e.,
$\frac{\partial \mathbf{b}_k[1]}{\partial \mathbf{p}} =\boldsymbol{h}(\mathbf{p};\boldsymbol{e}_1) , \frac{\partial \mathbf{b}_k[2]}{\partial \mathbf{p}} =\boldsymbol{h}(\mathbf{p};\boldsymbol{e}_2),  \frac{\partial \mathbf{b}_k[3]}{\partial \mathbf{p}} =\boldsymbol{h}(\mathbf{p};\boldsymbol{e}_3)$,
$ \boldsymbol{h}(\mathbf{p};\boldsymbol{e} )= \frac{2d_{r}}{\lambda} \left(  
{\boldsymbol{f} }(\mathbf{p}; \boldsymbol{e}, \mathbf{p}_{\rm TX}) +  {\boldsymbol{f} }(\mathbf{p}; \boldsymbol{e}, \mathbf{p}_{{\rm RX},k})  \right), 
$
and  $\boldsymbol{f}(\mathbf{p}; \boldsymbol{e}, \mathbf{p}_{k})  $ being defined in \eqref{eq-deri}. On the other hand, the partial derivative of $\eta_k^a$ and $\eta_k^e$ w.r.t. the axis angles $\psi_x$, $\psi_y$, and $\psi_z$ are given by $\frac{\partial \eta_k^a(\mathbf{p}, \mathbf{Q})}{\partial \psi_{i} } = \mathbf{b}_k^T (\mathbf{p}_{}) \frac{\partial \mathbf{q}_1 }{\partial \psi_{i}}$, and $
\frac{\partial \eta_k^e(\mathbf{p}, \mathbf{Q})}{\partial \psi_{i}} =  \mathbf{b}_k^T (\mathbf{p}_{}) \frac{\partial \mathbf{q}_2 }{\partial \psi_{i}} ,
 i={x,y,z}$, respectively, where $\mathbf{q}_1$ and $\mathbf{q}_2$ are the first and second columns of the matrix $\mathbf{Q}$, respectively. Based on the above, the Jacobian matrix $\nabla_{\boldsymbol{s}} \boldsymbol{\Gamma}$ can be obtained.

\section{Calculation of Jacobian Matrix $\nabla_{\boldsymbol{s}} \boldsymbol{\mathbf{\bar q}}$}
\label{AppendixC} 
The $(i,j)$-th element of $\nabla_{\boldsymbol{s}} \mathbf{\bar q}$ is given by $\frac{\partial \bar {\rm q} [i]}{ \partial  s[j]}$, where $\bar {\rm q}[i]$ is the $i$-th entry of the vector $\mathbf{ \bar q} = {\rm vec} (\mathbf{Q})$ and $s[j]$ is the $j$-th entry of the vector $\mathbf{s} =[ \mathbf{p}_{\rm I}^T, \boldsymbol{\psi}^T]^T$. As shown in \eqref{eq-Qseq}, the rotation matrix $\mathbf{Q}$ is regardless of the IRS location $\mathbf{p}_{\rm I}$. Hence, we only need to derive its partial derivative w.r.t. $\boldsymbol{\psi}$. For example, the first entry of $\mathbf{\bar q}$ is given by
$
\bar q[1] = \cos \psi_y\cos \psi_z,
$ and its partial derivative w.r.t. $\boldsymbol{\psi}$ can be calculated as $\frac{\partial \bar q[1]}{ \partial \boldsymbol{\psi}}=[0, - \sin \psi_y \cos \psi_z,  - \cos \psi_y \sin\psi_z]^T$. Similarly, the other entries of $\nabla_{\mathbf{s}} \mathbf{\bar q}$ can be calculated, for which the details are omitted for brevity.

\bibliography{newbib}
\bibliographystyle{IEEEtran}

\end{document}